# Cyber-human social co-operating system: Application design and implementation

Short title: Social Co-OS application

Takeshi Kato[1*], Misa Owa[2], Jyunichi Miyakoshi[2], Yasuhiro Asa[2], Takashi Numata[3], Yasuyuki Kudo[2], Tadayuki Matsumura[4], Kanako Esaki[5], Ryuji Mine[6], Toru Yasumura[7], Hikaru Matsunaka[7], Satomi Hori[7]

[1] Hitachi Kyoto University Laboratory, Institutional Advancement and Communications, Kyoto University, Kyoto, Japan
[2] Systems Innovation Center, Research & Development Group, Hitachi Ltd., Tokyo, Japan
[3] Sensing Integration Innovation Center, Research & Development Group, Hitachi Ltd., Tokyo, Japan
[4] Advanced Artificial Intelligence Innovation Center, Research & Development Group, Hitachi Ltd., Tokyo, Japan
[5] Next Research, Research & Development Group, Hitachi Ltd., Tokyo, Japan
[6] AI Transformation Promotion Project, Research & Development Group, Hitachi Ltd., Tokyo, Japan
[7] Design Center, Research & Development Group, Hitachi Ltd., Tokyo, Japan

* Correspondence
Email: kato.takeshi.3u@kyoto-u.ac.jp
ORCiD: https://orcid.org/0000-0002-6744-8606

CONFLICT OF INTEREST: The authors declare no conflict of interest.

FUNDING STATEMENT: The authors received no specific financial support for this study.

DATA AVAILABILITY STATEMENT: The data that support the findings of this study are available from the corresponding author upon reasonable request.

# Abstract

Reducing wealth inequality and resource waste is a global challenge. A fundamental problem within the capitalist economy, put simply, lies in the enslavement of labor and the colonization of resources. To address these issues, movements promoting digital democracy and cooperative platforms have emerged as viable alternatives to traditional capitalist systems. From the perspective of integrating information systems with real-world social, environmental, and economic systems, Cyber-Physical Systems have been proposed for applications in Industry 5.0 and Society 5.0. One such CPS is the Social Co-OS (cyber-human social co-operating system). Social Co-OS is a co-operating system between cyber and human societies that conceptualizes the social system as a dynamic, circular structure composed of three layers: individual behavior, interindividual interaction, and institutional formation. Within this framework, the cyber system supports collective decision-making and individual cooperative behavior across these layers. The objective of this study is to define a novel application architecture based on the Social Co-OS concept, design a user interface flow, and implement it in practice. Specifically, we develop a social impact evaluator, a pluralistic policy simulator, and a consensus-building facilitator, which constitute the deliberative and political loop of Social Co-OS. Additionally, we implement a personality estimator and a behavior change promoter, which constitute the operational and administrative loop, along with a common mediator that serves as the cyber-human interface. Through these implementations, we demonstrate that Social Co-OS applications can effectively support human social systems and offer practical utility for policy co-making and co-operation, as evidenced by examples grounded in real-world challenges. For future work, we aim to improve the Social Co-OS application by incorporating user feedback gathered from real-world field experiments. Furthermore, we aim to promote the wider adoption of digital democracy and cooperative platforms to address wealth inequality and resource waste, thereby contributing to the transformation of capitalist economies into more equitable and sustainable alternatives.

# 1. INTORODUCTION

Wealth inequality and resource waste have become major global social issues. According to the World Inequality Report 2022 [1], the top 10% of global wealth holders possess 76% of the world's wealth and 52% of its income, while accounting for 48% of global carbon emissions. The Global Resources Outlook 2024 [2] further indicates that consumption by the wealthy continues to drive greater resource use, thereby exacerbating both environmental degradation and social inequality.

Capitalist economies have expanded through the enclosure of the commons, resulting in the enslavement of labor and the colonization of resources [3, 4]. While global GDP continues to grow, this growth primarily benefits the super-wealthy—namely capitalists and corporate executives—while simultaneously increasing the consumption of resources and energy [5]. The widening gap of inequality undermines well-being, and excessive resource consumption drives environmental degradation. To reduce wealth inequality and resource consumption, while enhancing well-being and restoring natural ecosystems, viable alternatives to the capitalist economies are urgently needed [6].

Modifications to contemporary capitalism, such as stakeholder capitalism [7], progressive capitalism [8], and ethical capitalism [9], have been proposed. Simultaneously, alternatives to capitalism, such as the moral economy [10] and community economy [11, 12], have also been suggested. These proposals generally share several common elements: income restrictions and progressive taxation on the wealthy, redistribution of capital stock to address inequality, investment in public goods, and the revitalization of regional communities. In essence, they aim to protect democracy from the influence of capitalists and corporate executives in government, and to unenclose the commons, thereby enabling laborers can autonomously manage labor, resources, and capital.

Digital democracy [13, 14] and cooperative platforms [15, 16] are recognized as movements that seek to create alternatives to capitalism through the use of information technology. Digital democracy aims to foster a democratic and sustainable society by incorporating diverse opinions through active citizen participation. Cooperative platforms pursue a fair and sustainable economy by enabling laborers and consumers to jointly own and autonomously operate digital platforms. Collectively, these initiatives can be understood as efforts to establish new socio-economic systems through democratic decision-making and collaborative operation within the digital society.

From the perspective of integrating information systems with real-world social, environmental, and economic systems, approaches emphasizing human-centeredness and sustainability through Cyber-Physical Systems (CPS) are referred to as Industry 5.0 [17] and Society 5.0 [18]. While the body of previous research on these topics is extensive, a notable example is a review paper on Human-Cyber-Physical Systems (HCPS) in Industry 5.0 [19], which introduces three paradigms: Human-in-the-Loop (HitL), which integrates human perception into manufacturing systems; Human-on-the-Loop (HotL), which incorporates human decision-making onto manufacturing systems; and Human-in-the-Society (HitS), which connects human social networks with industrial ecosystems. Furthermore, a review paper on CPS supporting a sustainable circular economy (CE) [20] provides specific examples of CPS applications across the six stages of the CE: procurement, design, manufacturing, distribution, use, and recovery.

While previous studies have examined Industry 5.0 and the CE, there is also a review paper [21] that focuses on human behavior in both virtual and real spaces. This paper, situated within an interdisciplinary field spanning psychology, sociology, and computer science, introduces topics such as the relationship between digital and face-to-face communication [22], the influence of avatars in virtual and augmented reality environments [23], the impact of digital communication on human social life [24], and the effects of Social Co-OS (cyber-human social co-operating system)—which facilitates collaboration between cyber and human societies—on well-being [25].

Here, Social Co-OS is a co-operating system between cyber and human societies that conceptualizes the social system as a dynamic, circular structure consisting of three layers: individual behavior, interindividual interaction, and institutional formation [25]. Within this framework, the cyber system supports collective decision-making and individual cooperative behavior across these layers. This system is not limited to manufacturing contexts; rather, it resembles HitL in its circulation between cyber systems and groups or individuals, HotL in its integration of cyber systems with human decision-making, and HitS in its connection between cyber systems and human society. Furthermore, when applied to the CE, Social Co-OS supports decision-making and cooperative behavior at each stage of the CE lifecycle. The inequality and environmental issues previously discussed ultimately stem from human behavior and social structures. Digital democracy and cooperative platforms, which aim to provide alternatives to capitalism, engage in social decision-making and cooperative behavior. Therefore, this study focuses on Social Co-OS, viewing it as a specific type of CPS relevant to addressing these social issues.

The concept of Social Co-OS is grounded in the following theoretical frameworks: According to Luhmann's social systems theory [26], a social system is an autopoietic system composed of circular communication; according to Aoki's social institutional theory [27], a social system is a circular structure in which individual dispositions trigger actions, and these actions generate recursive states, collectively forming social institutions, which in turn influence individual dispositions; according to Heath's social norms theory [28], social norms emerge from mutual expectations and recognition within interindividual interactions, thereby shaping the order and institutions of social systems; according to Kahneman's dual-process theory [29], human information processing consists of a fast, automatic, and intuitive system as well as a slow, deliberative, and inferential system; according to Wilson's politics-administration dichotomy [30], social systems comprise fast operations and interventions in administration, and slow deliberation and institutional formation in politics.

Thus, Social Co-OS is founded on fundamental principles of human and social science. These principles provide the theoretical foundation for social institutions, norms, decision-making, and operations, with the aim of addressing the problems of wealth inequality and resource waste. Furthermore, Social Co-OS posits that the paradigm in Society 5.0 shifts away from conventional models—such as physics models (data analysis → retrospective theoretical explanation → prediction) and historical models (historical analysis → retrospective future insight → prevention)—toward a real-time and dynamic clinical medicine model (diagnostics → prognostic prediction → intervention) [25]. In other words, Social Co-OS can be described as a system that offers prescriptions for social issues based on a clinical medicine model. However, in practice, previous research has only presented the concept and basic functionalities of Social Co-OS.

This study aims to support real-world human social activities by building upon the concept of Social Co-OS as a CPS, defining its core functions as applications, specifying a novel architecture, and designing it as a general-purpose tool. Furthermore, to address the issues of wealth inequality and resource waste, the study seeks to promote the wider adoption of digital democracy and cooperative platforms, thereby contributing to the transformation toward viable alternatives to capitalist economies.

The remainder of this paper is structured as follows: In the Methods section, we first present the system and application architecture of Social Co-OS, and describe the specific user interface flow as part of the application function design; next, in the Results section, we provide actual user interface screens as implementation outcomes of the application functions, and demonstrate their usefulness through examples simulating real-world challenges; in the Discussion section, we revisit the achievements of Social Co-OS as an application tool, identify future technical challenges, and outline broader prospects for alternatives to capitalist economies; in the Conclusion section, we briefly summarize the content presented in the preceding sections.

## 2. METHODS

### 2.1 Social Co-OS architecture

Social Co-OS, as described above, is a cyber-human social co-operating system [25]. The system architecture of Social Co-OS is illustrated in Figure 1. Human society is conceptualized as a circular system comprising three layers: individual behavior, interindividual interaction, and institutional formation. The cyber system of Social Co-OS consists of a dual-loop structure involving both slow and fast processes. The slow loop functions as a deliberative and political process, providing diagnostics, prognostic predictions, and interventions for collective intentions, thereby supporting institutional formation—namely, consensus building. The fast loop serves as an operational and administrative process that diagnoses individual behavior and, based on prognostic insights, intervenes to promote interindividual interactions—i.e., cooperative behavior. Notably, "co-operating" carries a dual meaning: it refers both to the integration between cyber and human societies, and to the interplay between the slow and fast loops.

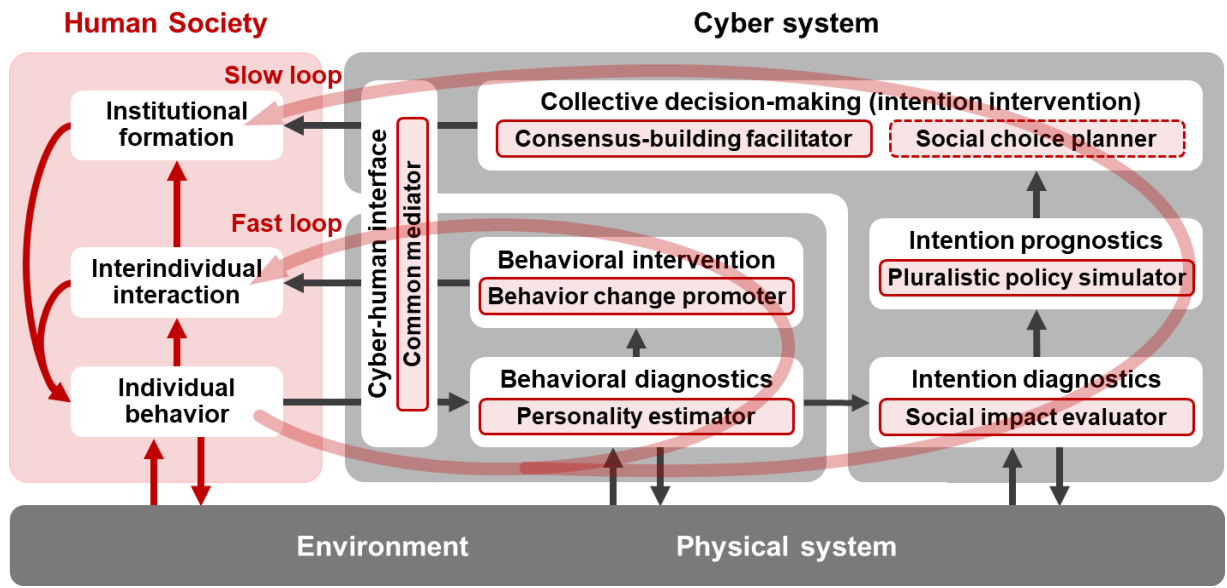

**FIGURE 1.** Social Co-OS system architecture.

In this study, we define each core function as an application based on the Social Co-OS system architecture shown in Figure 1. The details are as follows:

**Slow Loop:**
- **Collective Intention Diagnostics:** The social impact evaluator collects individuals' subjective intentions regarding social issues and constructs a logic model to quantitatively represent the influence of policies on social impact. A logic model is a structured logic tree that illustrates the sequence of inputs → activities → outputs → outcomes → impacts, clarifying the relationship between the policies (inputs) and the resulting impact they bring about.
- **Collective Intention Prognostics:** Based on the policies outlined in the social impact evaluator, the pluralistic policy simulator constructs a multi-agent model representing the policies and performs numerical simulations to quantitatively present objective prognostic effects. Policy variations are defined as parameters within the multi-agent model, and diverse policy scenarios are generated by combining these parameters.
- **Collective Decision-Making:** From the diverse policy scenarios generated by the pluralistic policy simulator, the social choice planner extracts several representative choices (this extraction step may be omitted if the choices are self-evident). The consensus-building facilitator then assists all participants in reaching a consensus on a particular choice. Here, consensus building is defined as "a process of compromise and synthesis meant to produce decisions that no one finds so violently objectionable that they are not willing to at least assent" [31]. During facilitation, based on the representative choices, participants are presented with three types of proposals: a permissible proposal, a compromise proposal, and a sublated proposal, with the aim of achieving final consensus.

**Fast Loop:**
- **Individual Behavioral Diagnostics:** When implementing a policy that has reached consensus through collective decision-making, the personality estimator assesses the personalities of subjects targeted by the policy, as these are important indicators of cooperative behavior. Estimation methods include online surveys and behavioral observation, using indicators such as Social Value Orientation (SVO) and Big Five traits. This study employs online surveys and SVO as the primary estimation methods.

- **Individual Behavioral Intervention:** The behavior change promoter utilizes a social-psychological model trained through machine learning on data from numerous psychological experiments involving social dilemma problems—situations in which individual interests conflict with collective ones. This model predicts the prognostic effects of interventions aimed at promoting cooperative behavior in policy operation and suggests appropriate intervention measures. The content of the consensus policy and the individual SVO, as estimated by the personality estimator, are set as input parameters for the social-psychological model.

**Interface:**
- **Cyber-Human Interface:** Within the common mediator, avatars facilitate communication between cyber systems and human users to ensure the smooth execution of both the slow and fast loops. For instance, the common mediator can assist with the following tasks: constructing logic models in the social impact evaluator, generating multi-agent models in the pluralistic policy simulator, facilitating consensus in the consensus-building facilitator, administering online surveys in the personality estimator, and implementing interventions in the behavior change promoter. This study primarily focuses on the facilitation function.

Based on the above, Figure 2 presents an overview of the user interface flow, illustrating the application architecture of Social Co-OS. Users first access the login screen, then select an application from the menu screen: social impact evaluator, pluralistic policy simulator, consensus-building facilitator, personality estimator, or behavior change promoter. The common mediator facilitates cyber-human communication throughout the entire user interface.

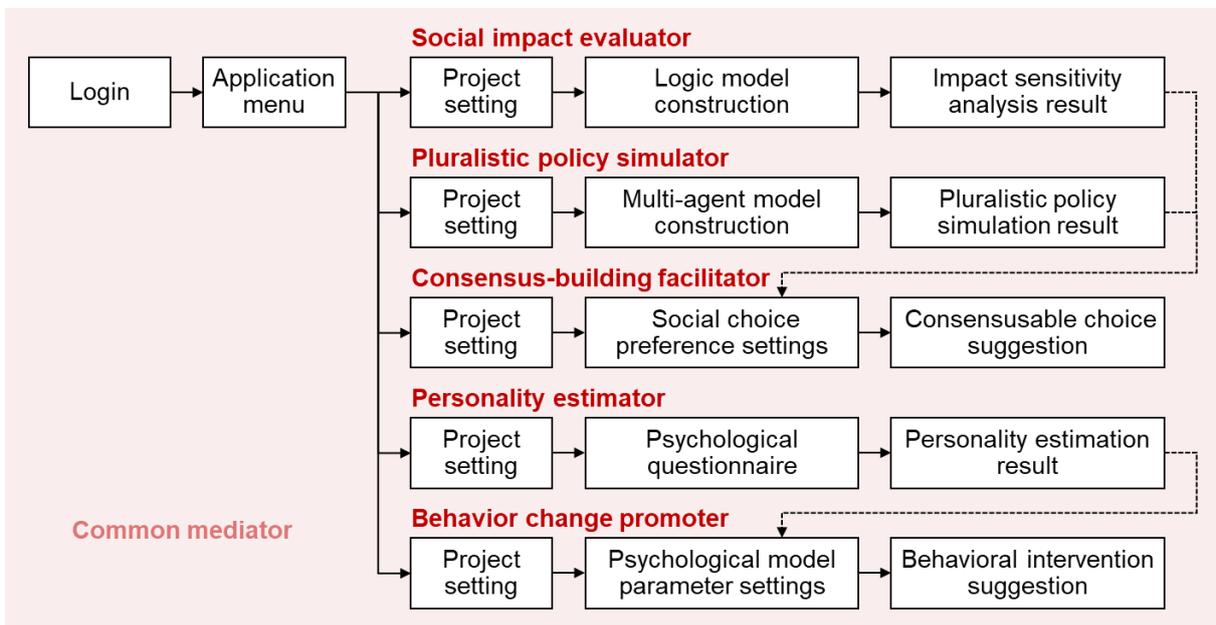

**FIGURE 2.** Social Co-OS application architecture: User interface flow diagram.

For each application, users first configure policy projects related to social issues via the project settings screen. This setup enables multiple applications to be operated in a coordinated manner for a common policy project, or conversely, to be operated independently for different policy projects. The user interface following project setup is then configured as follows:

- **Social Impact Evaluator:** First, deliberation participants construct a subjective logic model. They input elements (inputs, activities, outputs, outcomes, and impacts), connect these elements using arrows (relationships), and then assign weights to each connection. Next, based on the constructed logic model, the evaluator analyzes the sensitivity (effect) of the impact (goal) to the inputs (policies), and outputs the calculation results.
- **Pluralistic Policy Simulator:** First, deliberation participants construct an objective multi-agent model. They input agents (inputs representing policies, intermediate outcomes, and social, environmental, and economic values), and connect multiple inputs to ternary values via intermediate outcomes using arrows (relationships). Next, participants set parameters for each input. The simulator then analyzes the sensitivity of the ternary values to the inputs (policy elements), and outputs the calculation results. By varying the combinations of parameters, multiple policy scenarios can be compared.

- **Consensus-Building Facilitator:** First, deliberation participants receive policy choices from the social impact evaluator or pluralistic policy simulator and specify their preferences for these choices. Specifically, they input their preference order, permissible ranges, and factor evaluations for multiple choices. In factor evaluation, the importance of various factors constituting each policy is assessed. Next, the facilitator presents participants with three types of consensusable proposals: a permissible proposal, a compromise proposal, and a sublated proposal, generated through the processes of permissible meeting analysis, compromise choice exploration, and sublated choice creation, respectively. The facilitator then aims to reach consensus through dialogue with the participants.
- **Personality Estimator:** First, subjects involved in the operation of a consensus policy take a psychological questionnaire. The subjects answer multiple questions regarding the psychological indicator SVO. Next, based on their responses, the estimator determines their orientation (altruistic, prosocial, individualistic, or competitive) and outputs the corresponding results.
- **Behavior Change Promoter:** First, the policy operator configures feature parameters for the machine-learning-trained social-psychological model. These parameters represent factors related to policy operation, such as the structure of social dilemmas, the attributes of subjects, the relationships between subjects, and the presence or absence of rewards and punishments. The personality, as determined by the personality estimator, is reflected in the subject's attributes. Next, the behavior change promoter uses the configured social-psychological model to predict the prognostic effects of multiple interventions aimed at encouraging cooperative behavior. It then presents recommended interventions aligned with the policy objectives in ranked order.
- **Common Mediator:** Avatars are employed across the five applications mentioned above to facilitate communication with human users. A variety of avatar motions are available to accommodate diverse purposes and usage contexts. Deliberation conveners and policy operators can select appropriate avatar motions according to the specific requirements of each application scenario.

Thus far, we have presented the system architecture of Social Co-OS and a newly defined application architecture. In the subsequent sections, we detail each application: social impact evaluator, pluralistic policy simulator, consensus-building facilitator, personality estimator, behavior change promoter, and common mediator.

## 2.2 Social impact evaluator

In the social impact evaluator, deliberation participants construct a logic model to diagnose collective intentions and evaluate the social impact of policies. Figure 3 presents an overview of the user interface flow of the social impact evaluator.

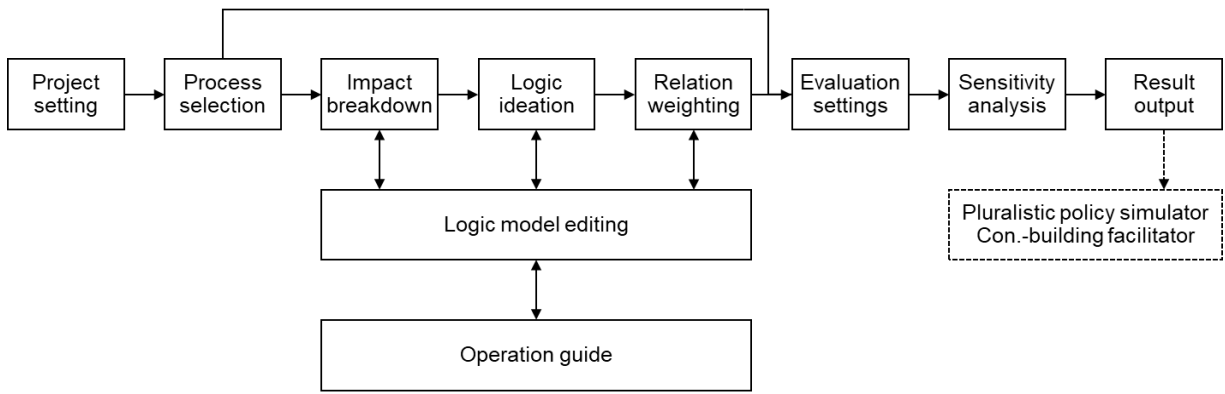

**FIGURE 3.** Social impact evaluator: User interface flow diagram.

The following list outlines the process step by step, in accordance with the flow illustrated in Figure 3.
- **Project Setting:** Create a new project or load an existing one.
- **Process Selection:** For new projects, proceed to the next step; for existing projects, proceed to the evaluation settings.
- **Logic Model Construction:** Break down the impact (policy goal), ideate logic elements (inputs, activities, outputs, outcomes), and subjectively assign weights to the relationships between elements while editing the logic model. An operation guide is displayed during the editing process.
- **Evaluation Settings:** Choose either simplified or advanced analysis. For advanced analysis, configure parameters such as policy frequency, expected effect, and expected attenuation.
- **Sensitivity Analysis:** Calculate the sensitivity (effect) of the impact (goal) to the inputs (policies), and display the results in graphs.
- **Result Output:** Export files containing the logic model and the sensitivity analysis results. If needed, transfer policy choices to the pluralistic policy simulator or consensus-building facilitator.

For further details on social impact evaluation, see Reference [32]. An overview is provided below.

The relationship between social impact and input is expressed as follows. Let social impact be denoted by $y_{SI}$, and input variables by $x_i$ $(i = 1, 2, \cdots, n)$. The logic model can be represented as a polynomial function $f_{LM}$ that links $y_{SI}$ and $x_i$, as shown in Equation (1). Intermediate logic elements are embedded within this polynomial function. The sensitivity $S_{SI_i}$ of the impact $y_{SI}$ to the input $x_i$ is defined as in Equation (2).

$$y_{SI} = f_{LM}(x_1, x_2, \cdots, x_i, \cdots, x_n) \tag{1}$$

$$S_{SI_i} = \frac{dy_{SI}}{dx_i} \tag{2}$$

Inputs $x_i$ with high sensitivity $S_{SI_i}$, or combinations thereof, exert a significant impact as collective intention diagnostics. Subsequently, policies subjectively identified as desirable by the social impact evaluator are adopted by the pluralistic policy simulator or the consensus-building facilitator.

## 2.3 Pluralistic policy simulator

In the pluralistic policy simulator, deliberation participants construct a multi-agent model to prognose collective intention and simulate the objective values of policies. Figure 4 presents an overview of the user interface flow of the pluralistic policy simulator.

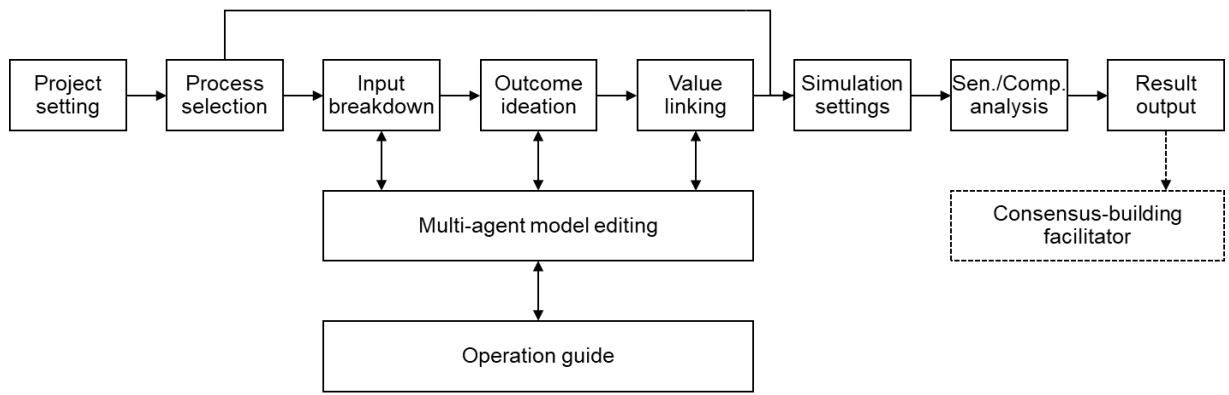

**FIGURE 4.** Pluralistic policy simulator: User interface flow diagram.

The following list outlines the process step by step, in accordance with the flow illustrated in Figure 4.
- **Project Setting:** Create a new project or load an existing one.
- **Process Selection:** For new projects, proceed to the next step; for existing projects, proceed to the

evaluation settings.
- **Multi-agent Model Construction:** Break down the inputs that constitute policies, ideate intermediate outcomes, and link them to social, environmental, and economic values while editing the multi-agent model. An operation guide is displayed during the editing process.
- **Simulation Settings:** Configure parameters for each of the multiple inputs (policy elements). A single policy is represented by a specific combination of input parameters.
- **Sensitivity/Comparative Analysis:** Compute the sensitivity of ternary values to inputs, visualize the results using graphs, and compare input parameter sets across different policies.
- **Result Output:** Export files containing the multi-agent model and the sensitivity/comparative analysis results. If needed, transfer policy choices to the consensus-building facilitator.

For further details on pluralistic policy simulation, see Reference [33]. An overview is provided below.

The relationship between ternary values and inputs is expressed as follows. Let ternary values be denoted by $y_{V_j}$ ($j = soc, env, eco$), and inputs by $x_i$ ($i = 1, 2, \cdots, n$). The multi-agent model can be represented as a polynomial function $f_{MM}$ that links $y_{V_j}$ and $x_i$, as shown in Equation (3). Intermediate outcomes are embedded within this polynomial function. Since a policy is defined as a set of inputs $x_i$, let policy $p_k$ ($k = 1, 2, \cdots, m$) be represented as $\{x_{1_k}, x_{2_k}, \cdots, x_{i_k}, \cdots, x_{n_k}\}$. The ternary value corresponding to policy $p_k$ is then expressed as $y_{V_{j_k}}$.

$$y_{V_j} = f_{MM}(x_1, x_2, \cdots, x_i, \cdots, x_n) \tag{3}$$

To visualize the relationship among the social value $y_{V_{soc_k}}$, environmental value $y_{V_{env_k}}$, and economic value $y_{V_{eco_k}}$ of policy $p_k$ in a ternary graph, these values are subjected to a two-step normalization process. First, for each of the three values, the maximum and minimum values of $y_{V_{j_k}}$ across policy $p_k$ are used to normalize them to $Y_{V_{soc_k}}$, $Y_{V_{env_k}}$, and $Y_{V_{eco_k}}$, as shown in Equation (4). Next, using the sum of the three normalized values, they are further normalized to $\Upsilon_{V_{soc_k}}$, $\Upsilon_{V_{env_k}}$, and $\Upsilon_{V_{eco_k}}$, as shown in Equation (5). These final values satisfy the condition $\Upsilon_{V_{soc_k}} + \Upsilon_{V_{env_k}} + \Upsilon_{V_{eco_k}} = 1$ on the ternary graph. The sensitivity $S_{V_{ji_k}}$ of the value $y_{V_{j_k}}$ to the input $x_{i_k}$ in policy $p_k$ is defined as in Equation (6).

$$Y_{V_{j_k}} = \frac{y_{V_{j_k}}}{\max_l y_{V_{j_l}} - \min_l y_{V_{j_l}}} \tag{4}$$

$$\Upsilon_{V_{j_k}} = \frac{Y_{V_{j_k}}}{Y_{V_{soc_k}} + Y_{V_{env_k}} + Y_{V_{eco_k}}} \tag{5}$$

$$S_{V_{ji_k}} = \frac{dy_{V_{j_k}}}{dx_{i_k}} \tag{6}$$

A policy $p_k$, composed of inputs $x_i$ with high sensitivity $S_{V_{ji_k}}$ and a balanced distribution of three values $\Upsilon_{V_{j_k}}$, demonstrates high effectiveness as collective intention prognostics. Subsequently, policies identified as desirable—both subjectively by the social impact evaluator and objectively by the pluralistic policy simulator—are transferred to the consensus-building facilitator for decision-making.

### 2.4 Consensus-building facilitator
In the consensus-building facilitator, deliberation participants input their preferences regarding policy choices. Based on analytical results, the facilitator presents conensusable choices and engages in dialogue with participants to reach a final agreement through collective decision-making. Figure 5 presents an overview of the user interface flow of the consensus-building facilitator.

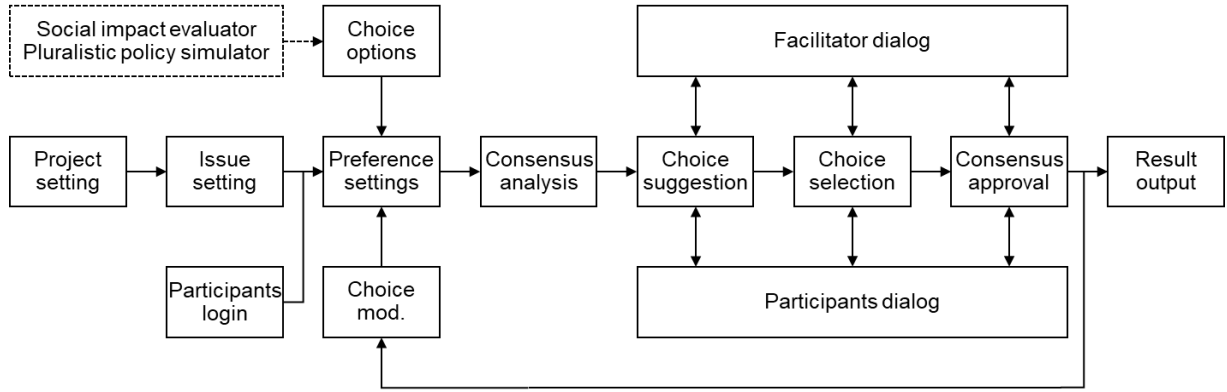

**FIGURE 5.** Consensus-building facilitator: User interface flow diagram.

The following list outlines the process step by step, in accordance with the flow illustrated in Figure 5.
- **Project Setting:** Create a new project or load an existing one.
- **Issue Setting:** Define the agenda for deliberation to determine policies addressing social issues. Concurrently, receive policy choice options related to the issues from both the social impact evaluator and the pluralistic policy simulator.
- **Preference Settings:** Deliberation participants input their preference order, permissible ranges, and factor evaluations for multiple policy choices. Permissible ranges are specified starting from the top-ranked preferences, and for factor evaluations participants assign importance to each factor constituting the policies.
- **Consensus Analysis:** Through permissible meeting analysis, compromise choice exploration, and sublated choice creation, derive a permissible proposal, a compromise proposal, and a sublated proposal, respectively. These are presented as consensusable choices.
- **Facilitation:** While engaging in dialogue with participants, the facilitator presents consensusable choices. Participants then discuss the feasible choices and indicate their approval or disapproval. This process is repeated iteratively. If consensus is not achieved, the facilitator modifies the choices and returns to the preference-setting process.
- **Result Output:** Export files containing preference data, consensus analysis results, and dialogue history.

For further details on permissible meeting analysis, compromise choice exploration, and sublated choice creation, see Reference [34]. An overview is provided below.

In permissible meeting analysis, let the preference order vector of participant $i$ be denoted by $x_i$ ($i = 1, 2, \cdots, n$), the permissible range within $x_i$ by $g_i$, and the widening operation applied to the permissible range by $f_{PMA}$. The permissible proposal $x_{PC}$ is defined as shown in Equation (7). The dimension of $x_i$ corresponds to the number of policy choices. The permissible proposal $x_{PC}$ is the one that requires the minimum number of widening operations $f_{PMA}$ across all $n$ participants.

$$x_{PC} = f_{PMA_{min}}(g_1(x_1), g_2(x_2), \cdots, g_i(x_i), \cdots, g_n(x_n)) \quad (7)$$

In compromise choice exploration, let the preference order vector of participant $i$ be denoted by $x_i$ ($i = 1, 2, \cdots, n$), and the replacement operation applied to the preference order be represented by $f_{CCE}$. The resulting vector $x_{CC}$, in which the $x_i$ vectors of all $n$ participants are aligned, is defined as shown in Equation (8). The vector $x_{CC}$ represents an arrangement in which the total number of replacement operations $f_{CCE}$ across all participants is as minimized as possible, and the distribution of replacements among participants is as balanced as possible. The compromise proposal $x_{CC_1}$ corresponds to the top-ranked choice in $x_{CC}$.

$$x_{CC} = f_{CCE_{min \cdot even}}(x_1, x_2, \cdots, x_i, \cdots, x_n) \quad (8)$$

In sublated choice creation, let the preference order vector of participant $i$ be denoted by $x_i$ ($i = 1, 2, \cdots, n$), the factor evaluation vector associated with $x_i$ be denoted by $h_i$, and the editing operation based on factor evaluation be represented by $f_{SCC}$. The sublated proposal $x_{SC}$ is defined as shown in Equation (9). The editing operation $f_{SCC}$ creates the sublated proposal $x_{SC}$ by incorporating, to the greatest extent possible, the content of higher-ranked choices and the factors deemed more important, while minimizing the inclusion of lower-ranked choices and less significant factors.

$$x_{SC} = f_{SCC}(h_1(x_1), h_2(x_2), \cdots, h_i(x_i), \cdots, h_n(x_n)) \tag{9}$$

The policy ultimately selected through the presentation of a permissible proposal $x_{PC}$, a compromise proposal $x_{CC_1}$, and a sublated proposal $x_{SC}$ represents a consensus reached via collective decision-making. To operate this policy in practice, the personality estimator and behavior change promoter are activated.

## 2.5 Personality estimator

In the personality estimator, the policy operator conducts a questionnaire survey targeting individual subjects to estimate personality traits associated with cooperative behavior, serving as individual behavioral diagnostics. Figure 6 presents an overview of the user interface flow of the personality estimator.

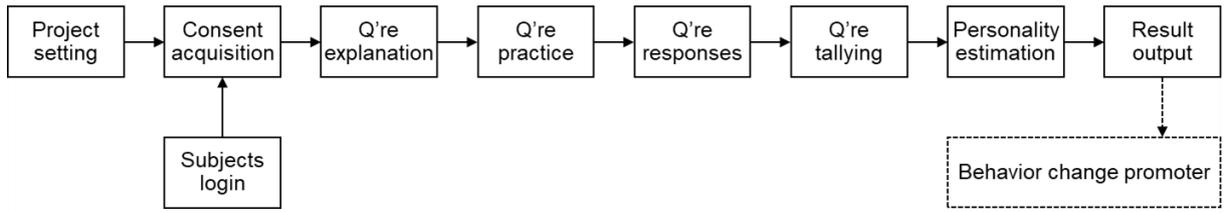

**FIGURE 6.** Personality estimator: User interface flow diagram.

The following list outlines the process step by step, in accordance with the flow illustrated in Figure 6.
- **Project Setting:** Create a new project or load an existing one.
- **Consent Acquisition:** Recruit subjects involved in policy operation and obtain their informed consent regarding participation in the questionnaire.
- **Questionnaire Explanation:** Provide a detailed explanation of the questionnaire contents.
- **Questionnaire Practice:** Conduct a practice session prior to administering the actual questionnaire.
- **Questionnaire Responses:** Collect responses from subjects for the actual questionnaire. This includes 15 items related to the psychological indicator known as SVO.
- **Questionnaire Tallying:** Aggregate and organize the questionnaire responses.
- **Personality Estimation:** Estimate the subjects' personality traits based on the aggregated results. This includes identifying the primary SVO type (altruistic, prosocial, individualistic, or competitive) and the secondary SVO (equality orientation) as a subsidiary trait.
- **Result Output:** Export files containing the questionnaire data and personality estimation results.
 For further details on personality estimation, see Reference [35, 36]. An overview is provided below.
 The relationship between questionnaire responses and personality traits is expressed as follows. Let the responses to the psychological questionnaire be denoted as $x_i\ (i = 1, 2, \cdots, n)$, and the tallying operation by $f_{PQ}$. The resulting personality vector $\boldsymbol{y_P}(y_{P_1}, y_{P_2})$ is defined as shown in Equation (10). The primary SVO is represented by the profit distribution between the subject and others, where $\boldsymbol{y_P}(y_{P_1}, y_{P_2})$ is a two-dimensional vector and $\boldsymbol{f_{PQ}}$ is a vector operation. The secondary SVO, reflecting the subject's equality orientation, is represented as a one-dimensional value.

$$\boldsymbol{y_P}(y_{P_1}, y_{P_2}) = \boldsymbol{f_{PQ}}(x_1, x_2, \cdots, x_i, \cdots, x_n) \tag{10}$$

The personality vector $\boldsymbol{y_P}(y_{P_1}, y_{P_2})$, derived from individual behavioral diagnostics, is transferred to the behavior change promoter as subject attributes for use in intervention design.

## 2.6 Behavior change promoter

In the behavior change promoter, the policy operator configures feature parameters related to policy operation within a social-psychological model, identifies interventions aimed at promoting cooperative behavior and their predicted effects, and subsequently applies these interventions to individual subjects. Figure 7 presents an overview of the user interface flow of the behavior change promoter.

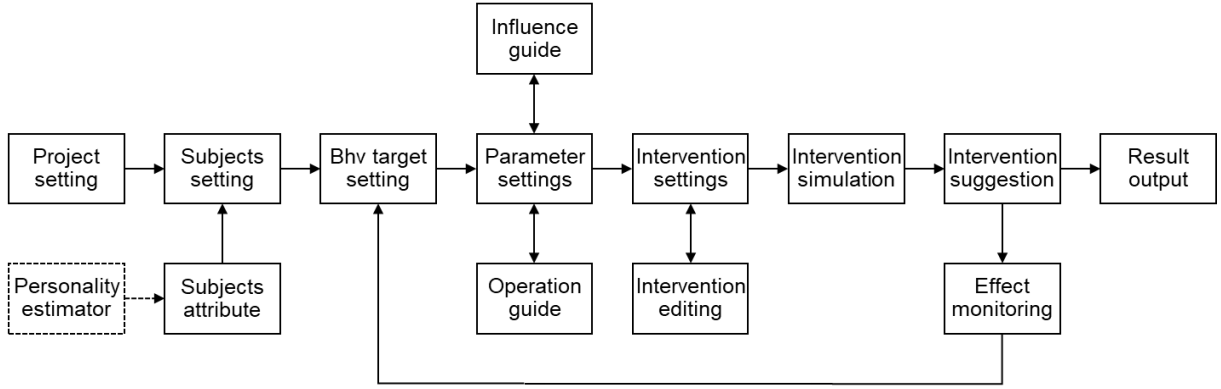

**FIGURE 7.** Behavior change promoter: User interface flow diagram.

The following list outlines the process step by step, in accordance with the flow illustrated in Figure 7.
- **Project Setting:** Create a new project or load an existing one.
- **Subject Setting:** Define the subjects for policy operation and import their attributes from the personality estimator.
- **Behavior Target Setting:** Specify the categories, characteristics, and frequency of issues related to policy operation.
- **Parameter Settings:** For the machine-learning-trained social-psychological model, configure initial parameters—such as the structure of the social dilemma, subject attributes, relationships among subjects, and the presence or absence of rewards and punishments—as features representing the policy-related issues. During setup, a feature influence guide and an operation guide are displayed.
- **Intervention Settings:** Define a new intervention or select multiple interventions from a predefined menu. An intervention is a parameter set that modifies the initial configuration and represents a psychological intervention targeting the subjects.
- **Intervention Simulation:** Simulate the effectiveness of multiple interventions in promoting cooperative behavior, and display the most desirable interventions in ranked order.
- **Intervention Suggestion:** For the selected intervention, present the expected effectiveness and sustainability, and apply the intervention to the subjects. Monitor its effectiveness; if it declines, return to the behavior target setting process.
- **Result Output:** Export files containing parameter configurations, intervention settings, and simulation results.

For further details on the social-psychological model, feature parameters, and interventions, see Reference [37]. An overview is provided below.

The relationship between the features of the psychological model and the cooperation rate is expressed as follows. Let the feature parameter be denoted by $x_i$ $(i = 1, 2, \cdots, n)$, and the psychological model by $f_{PM}$. The cooperation rate $y_{CR}$ is defined as shown in Equation (11). The model $f_{PM}$ is a neural network trained using machine learning on a large dataset of psychological experiment results. By specifying the behavioral target and initial parameters, the cooperation rate prior to intervention can be estimated. The sensitivity $S_{CR_i}$ of the cooperation rate $y_{CR}$ to the feature parameters $x_i$ is expressed as shown in Equation (12).

$$y_{CR} = f_{PM}(x_1, x_2, \cdots, x_i, \cdots, x_n) \tag{11}$$

$$S_{CR_i} = \frac{dy_{CR}}{dx_i} \qquad (12)$$

Feature parameters $x_i$ with high sensitivity $S_{CR_i}$, or combinations thereof, exhibit high effectiveness. By executing interventions that modify the initial parameters, the cooperation rate after intervention is improved compared to the rate prior to intervention. In other words, the behavior change promoter can facilitate cooperative behavior through individual behavioral intervention, thereby supporting policy operation.

Up to this point, we have explained the five applications of Social Co-OS. In the slow loop (deliberation/politics), the participants can formulate policies—that is, institutions—through the following components: the social impact evaluator (collective intention diagnostics), the pluralistic policy simulator (collective intention prognostics), and a consensus-building facilitator (collective decision-making). In the fast loop (operation/administration), the policy operator implements the policy by promoting cooperative behavior among subjects through the personality estimator (individual behavioral diagnostics) and the behavior change promoter (individual behavioral prognostics/intervention).

## 2.7 Common mediator

In the common mediator, avatars facilitate communication across the five components—social impact evaluator, pluralistic policy simulator, consensus-building facilitator, personality estimator, and behavior change promoter—in accordance with the objectives of deliberation conveners and policy operators, as well as the states of deliberation participants and policy subjects.

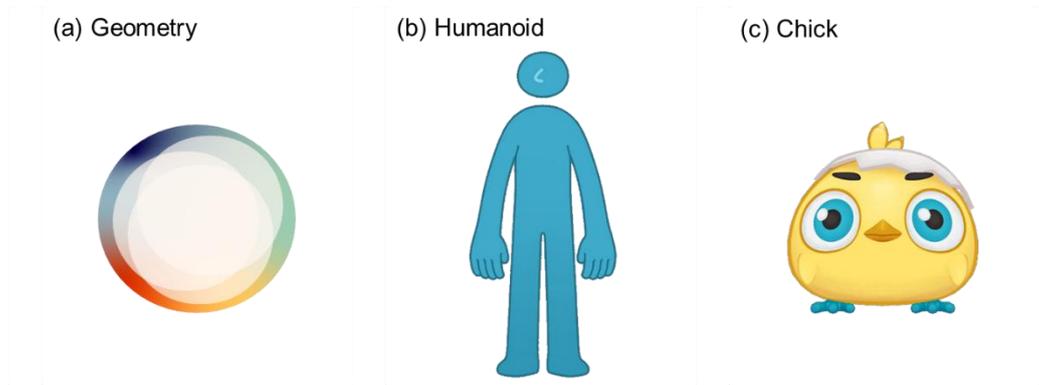

**FIGURE 8.** Common mediator: Three types of avatars.

The geometry-type avatar shown in Figure 8a conveys intentions and states by deforming its outer and inner circles and altering their colors. The humanoid-type avatar in Figure 8b expresses intentions and states through body movements, while the chick-type avatar in Figure 8c does so through facial expressions and body movements. These three types of avatars can be selected based on the agenda, policy content, participant attributes, and other contextual factors.

This study primarily generates motion data for avatars designed for use within the consensus-building facilitator. Table 1 presents the avatar motions corresponding to the facilitator, individual participants, and participant groups.

**TABLE 1.** Avatar motions in consensus-building facilitator.

| # | Facilitator | Individual participant | Participant group |
|---|---|---|---|
| 1 | Request | — | — |
| 2 | Introduction | — | — |
| 3 | Proposal | — | — |
| 4 | Neutral | Neutral | — |
| 5 | Divergence | Divergence | — |
| 6 | Convergence | Convergence | — |
| 7 | Confusion | Confusion | — |
| 8 | View change | View change | — |
| 9 | Cooperation | Cooperation | — |
| 10 | Ripe time | Ripe time | Ripe time |
| 11 | — | Approval | — |
| 12 | — | Opposition | — |
| 13 | — | — | Scattered |
| 14 | — | — | Division |
| 15 | — | — | Confrontation |
| 16 | — | Compromise | Consensus |
| 17 | — | Unrejection | Superficial agreement |

Facilitator motions: In Table 1, a single avatar representing the facilitator primarily conveys interactions with participants. For example, motions #5 and #6 encourage participants to diverge and converge their opinions, respectively; motion #8 prompts the expression of opinions from diverse perspectives; and motion #10 facilitates movement toward consensus once participants have exhausted their viewpoints and when the timing is ripe.

Participant motions: A single avatar represents the manner and behavioral state of an individual participant. For example, motions #5 and #6 respectively indicate that the participants' opinions are diverging or converging; motion #16 represents a compromise rather than full agreement; and motion #17 indicates that the participant is not rejecting the possibility of agreement.

Participant group motions: Multiple avatars represent the collective state of the participant group, in other words, the mood of the discussion. For example, motion #13 indicates that opinions are widely scattered; motion #14 reflects a division into majority and minority factions; motion #16 represents a state in which consensus is reached without any participant refusing to agree; and motion #17 depicts a superficial agreement lacking genuine intention.

Note that the avatar motions presented in Table 1 are primarily designed for use within the consensus-building facilitator, but they can be adapted for other applications. For example, facilitator motions may be utilized in various contexts, such as when the social impact evaluator encourages deliberation participants to construct a logic model; when the pluralistic policy simulator prompts participants to develop a multi-agent model; when the personality estimator guides policy subjects in completing a questionnaire; and when the behavioral change promoter implements psychological interventions for the subjects.

## 3. RESULTS

### 3.1 Social Co-OS opening

This section presents the implementation results of the user interface based on application functionality. First, Figure 9 illustrates the opening screens of the Social Co-OS user interface.

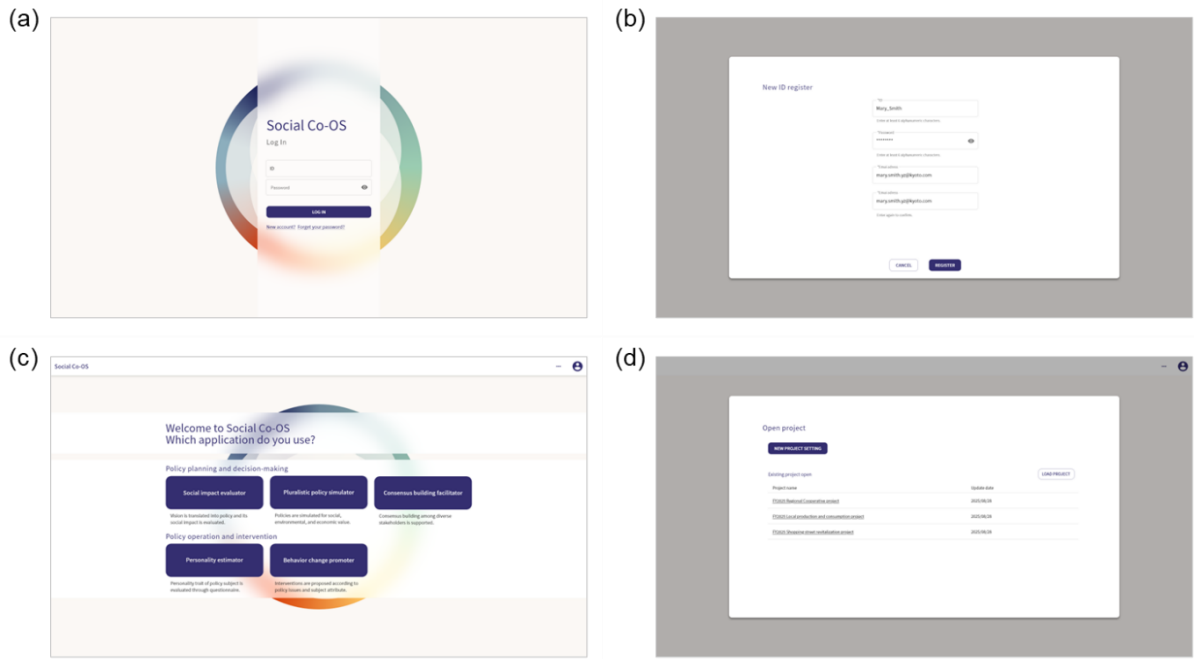

**FIGURE 9.** Social Co-OS: User interface opening screens.

Figure 9a shows the login screen (as previously shown in Figure 2). Figure 9b displays the ID registration screen. Figure 9c presents the menu selection screen (also referenced in Figure 2), where users select an application from the social impact evaluator, the pluralistic policy simulator, the consensus-building facilitator, the personality estimator, and the behavior change promoter.

Figure 9d shows the project setting screen for the five applications listed in Figure 2. For each application, the deliberation convener or policy operator configures the policy project. Users can create new projects, load templates, or access existing projects. Templates are provided as representative examples of regional issues, such as regional commons (e.g., resources, energy), municipal planning meetings, citizen participation councils, cooperative business management, and local production-for-local-consumption supply chains.

This study uses a policy project focused on the utilization of unused stocks in a regional community as a representative example. For this common case, the following subsections describe the user interfaces of the five applications. Unused stocks include abandoned farmland, neglected forests, and vacant homes, which—although privately owned—also function as commons for residents, serving as productive resources and natural environments [38]. While the emergence of unused stocks is largely attributed to population aging and declining birthrates, leaving them unaddressed constitutes a form of enclosure of the commons, as disscused in the Introduction. Policies are therefore needed to unenclose unused stocks to make them accessible to interested parties, promote their cooperative utilization as productive resources and natural environments, and reduce inequalities between owners and users, both of whom are residents. For further details on the composition of regional communities, refer to the context of Japanese rural areas [39] and Appendix A.1.

In the following subsections, we present the results of the five applications, focusing on the main policy choices: (A) production projects utilizing abandoned farmland and neglected forests; (B) regional energy projects leveraging abandoned farmland and neglected forests; (C) sales businesses utilizing regional products derived from farmland and forests, including abandoned areas; and (D) renovation and tourism businesses connecting vacant homes with immigrants and tourists.

## 3.2 Social impact evaluator

Figure 10 presents representative screens from the user interface of the social impact evaluator.

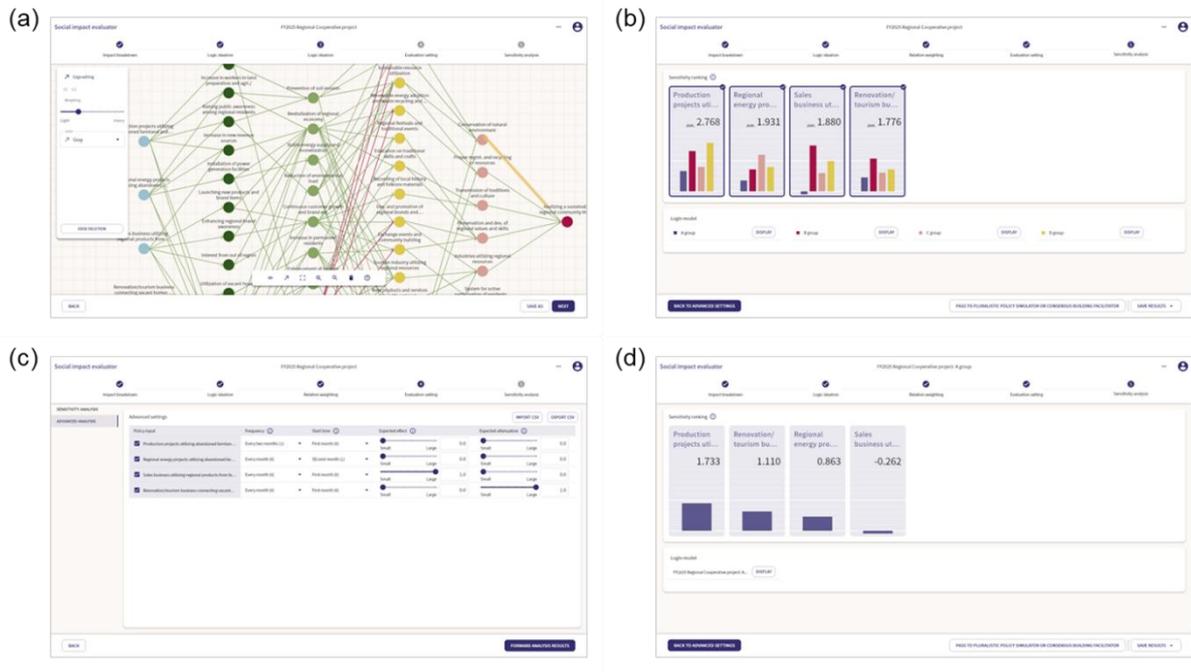

**FIGURE 10.** Social impact evaluator: User interface screens.

Figure 10a shows the logic model editing screen (as referenced in Figure 3). In the policy project, deliberation participants first define the right-hand side of the logic model (policy goal) by breaking down the intended social impact. They then develop the logic from left to right by ideating inputs (policies), activities, outputs, and outcomes, followed by assigning weights to the edges that represent relationships between logic elements. The completed logic model corresponds to Equation (1). For further details on the logic model, see Appendix A.2.

Figure 10b displays the sensitivity analysis screen (also referenced in Figure 3). Using Equation (2), the sensitivity of the impact to each input is calculated, and the inputs are ranked accordingly. The bar charts represent sensitivity values, with different colors indicating comparisons of results obtained from logic models constructed by multiple groups working on the same policy project.

Figure 10c shows the evaluation settings screen for advanced analysis. Using sliders, the frequency of the policy, the start date, the expected effect, and the expected attenuation are configured. Figure 10d presents the advanced analysis screen based on these settings. It displays the results of the advanced analysis for the logic model defined by a specific group. In the example project, this group is envisioned as comprising young and middle-aged members engaged in agriculture in mountainous and rural areas, who are natives of the region, and who possess a strong sense of community.

The results in Figure 10d indicate that, in the unused stock utilization project featured in the example, regarding the social issue of utilizing commons, the policy sensitivity (effectiveness) ranks as follows: (C) sales businesses utilizing regional products from farmland and forests, (B) regional energy projects utilizing abandoned farmland and neglected forests, (A) production projects utilizing abandoned farmland and neglected forests, and (D) renovation and tourism businesses connecting vacant homes with immigrants and tourists. This is likely because the group consists of agricultural workers native to the region.

In contrast to these subjective social impact evaluation results, the pluralistic policy simulator in the next subsection objectively examines policies for unused stock utilization. Note that the results in Figure 10 depend on the structure of the logic model and the edge weights; therefore, this outcome represents only one example, and the same applies to the results in the following subsections.

### 3.3 Pluralistic policy simulator
Figure 11 shows representative screens of the pluralistic policy simulator's user interface.

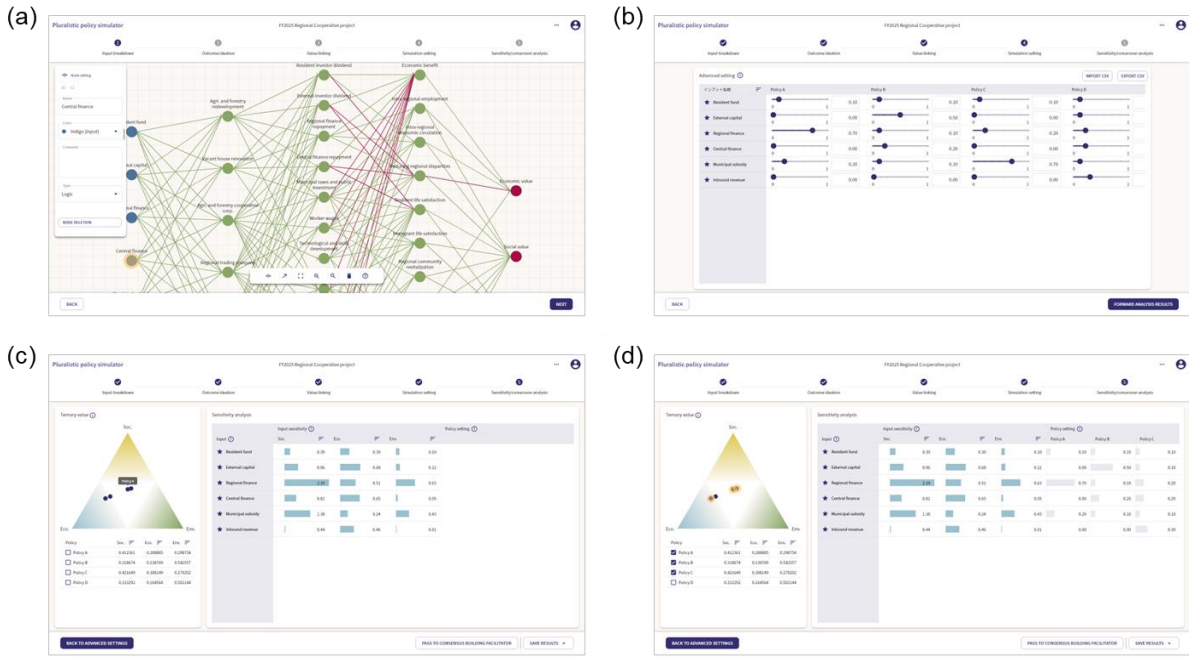

**FIGURE 11.** Pluralistic policy simulator: User interface screens.

Figure 11a displays the editing screen for the multi-agent model shown in Figure 4. In the policy project, deliberation participants break down the input (policy elements) on the left side of the model, ideate intermediate outcomes, and link them to the social, environmental, and economic values on the right side.

Figure 11b presents the simulation settings screen as shown in Figure 4. For each of the multiple policies, a combination of input parameters is configured using sliders. The completed multi-agent model for one policy corresponds to Equation (3). For details on the multi-agent model, see Appendix A.3.

Figure 11c shows the sensitivity analysis screen as shown in Figure 4. For a single policy, the social, environmental, and economic values calculated using Equation (5) are plotted on a ternary graph, and the results of the sensitivity analysis for each social, environmental, and economic value with respect to each input are displayed in a bar graph using Equation (6). The multiple plots in the ternary graph represent multiple policies, and selecting one plot displays the sensitivity analysis results for that policy, i.e., the sensitivity to each of the multiple policy elements constituting a single policy.

Figure 11d presents the sensitivity/comparison analysis screen as shown in Figure 4. When multiple plots are selected on the ternary graph in addition to the plot selected in Figure 11c, the input parameters for those policies are displayed to the right of the sensitivity analysis results. By comparing the parameter sets of multiple policies together with the sensitivity analysis results for a single policy, it is possible to determine the plot position on the ternary graph—i.e., the effect of policy elements on social, environmental, and economic values—and to identify what combination of policy elements is desirable for achieving the policy goal.

The results in Figure 10d indicate that, in the unused stock utilization project, (B) regional energy projects utilizing abandoned farmland and neglected forests and (D) renovation and tourism businesses connecting vacant homes with immigrants and tourists exhibit closer economic values, whereas (A) production projects utilizing abandoned farmland and neglected forests and (C) sales businesses utilizing regional products from farmland and forests demonstrate higher social and environmental values. The sensitivity analysis results show that external capital and central finance contribute primarily to economic value, while regional finance and municipal subsidies contribute to social and environmental values. These findings suggest that (A) and (C), which rely on regional residents and regional capital, offer a better balance of social, environmental, and economic values than (B), which involves capital from outside the region, or (D), which depends on immigrants and tourists.

When combined with the results of the social impact evaluator in the previous subsection, the following policies are identified as choices for decision-making in the consensus-building facilitator discussed in the next subsection: (A) production projects, (B) energy projects, and (C) sales projects—excluding (D) renovation and tourism projects—as well as combinations of these policies: (A × B), (B × C), and (C × A).

### 3.4 Consensus-building facilitator

Figure 12 shows representative screens of the consensus-building facilitator's user interface.

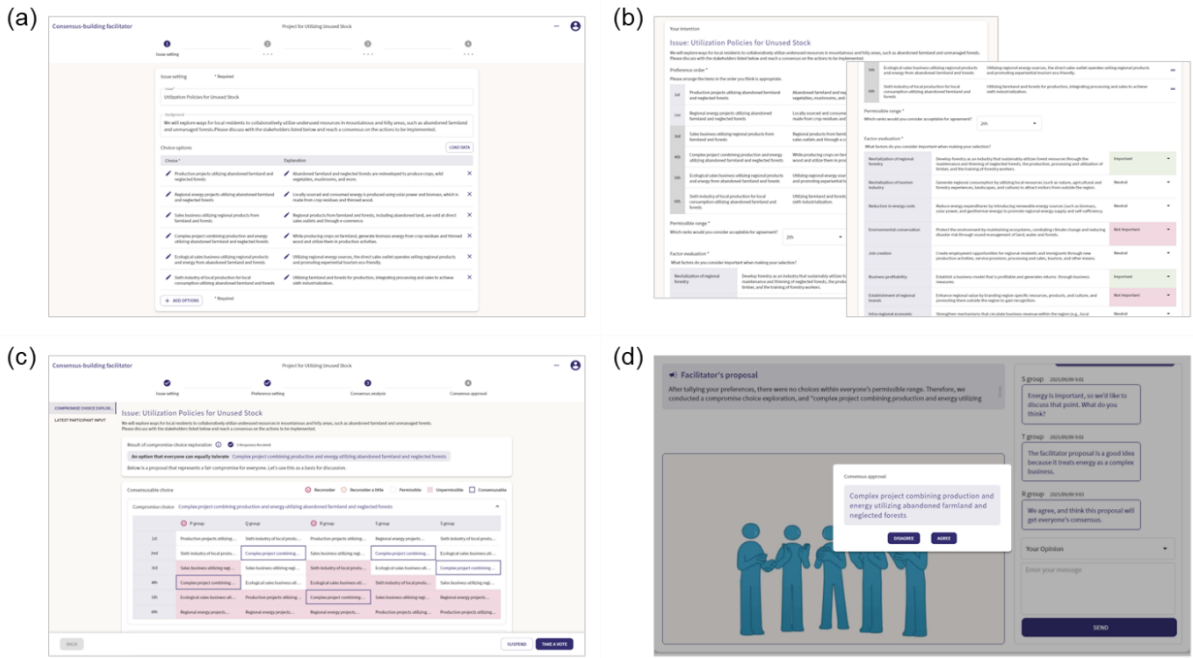

**FIGURE 12.** Consensus-building facilitator: User interface screens.

Figure 12a displays the issue-setting screen as shown in Figure 5. The deliberation convener sets the agenda and, based on the results of the social impact evaluator and pluralistic policy simulator, specifies six policy choices on this screen, while policy factors are configured in a scrollable section at the bottom, which is not visible here.

Figure 12b presents the preference-setting screen as shown in Figure 5. For six policy choices, the deliberation participants input their preference order and permissible range on the left panel, and their factor evaluation on the right panel. Of the three inputs, the first and second are used for permissible meeting analysis, the first for compromise choice exploration, and the third for sublated choice creation. Based on these inputs, consensusable choices are derived through three analysis processes using Equations (7), (8), and (9), respectively. For details on policy choices and policy factors, see Appendix A.4.

Here, five stakeholder groups are envisioned as follows: (P) young and middle-aged members engaged in agriculture in mountainous and rural areas, who are natives of the region and possess a strong sense of community; (Q) community members of the child-rearing generation who have relocated from urban areas as a countermeasure to the declining birthrate; (R) elderly regional members who have retired from agriculture and forestry due to aging; (S) members of private companies (e.g., renewable energy companies, distribution and sales companies, tourism operators); and (T) local government officials (e.g., general policy divisions, regional development divisions).

Figure 12c shows the screen displaying the consensus analysis results as shown in Figure 5. Here, the outcome of the compromise choice exploration is presented. For each group, a compromise proposal indicated by a blue frame is presented as a consensusable choice, and each group member is asked to reconsider their preference order. During facilitation, the facilitator—through dialogue with participants and the avatar mediation described in subsection 3.7—encourages the reselection of choices and aims to achieve final consensus.

Figure 12d presents the consensus approval screen as shown in Figure 5. Once the discussion is exhausted and the timing is ripe, the facilitator presents a single choice to the participants and seeks their approval for consensus. If approval (agreement that does not amount to rejection) is obtained from all participants, consensus is reached.

In Figure 12, the decision is made to select (A × B) a combined production and energy project utilizing abandoned farmland and neglected forests in the unused stock utilization project. This policy combines the second-highest sensitivity (B) and the third-highest sensitivity (A) for group (P) in Figure 10, and integrates the relatively high economic value (B) with the relatively high social and environmental values (A) shown in the ternary graph in Figure 11. This indicates that diverse stakeholders have reached a consensus to adopt a policy that strikes a good balance among social, environmental, and economic values, while leaning slightly toward economic value. In the following subsections, we shift our focus to the personality estimator and behavior change promoter for operating the (A × B) policy.

### 3.5 Personality estimator

Figure 13 shows representative screens of the personality estimator's user interface.

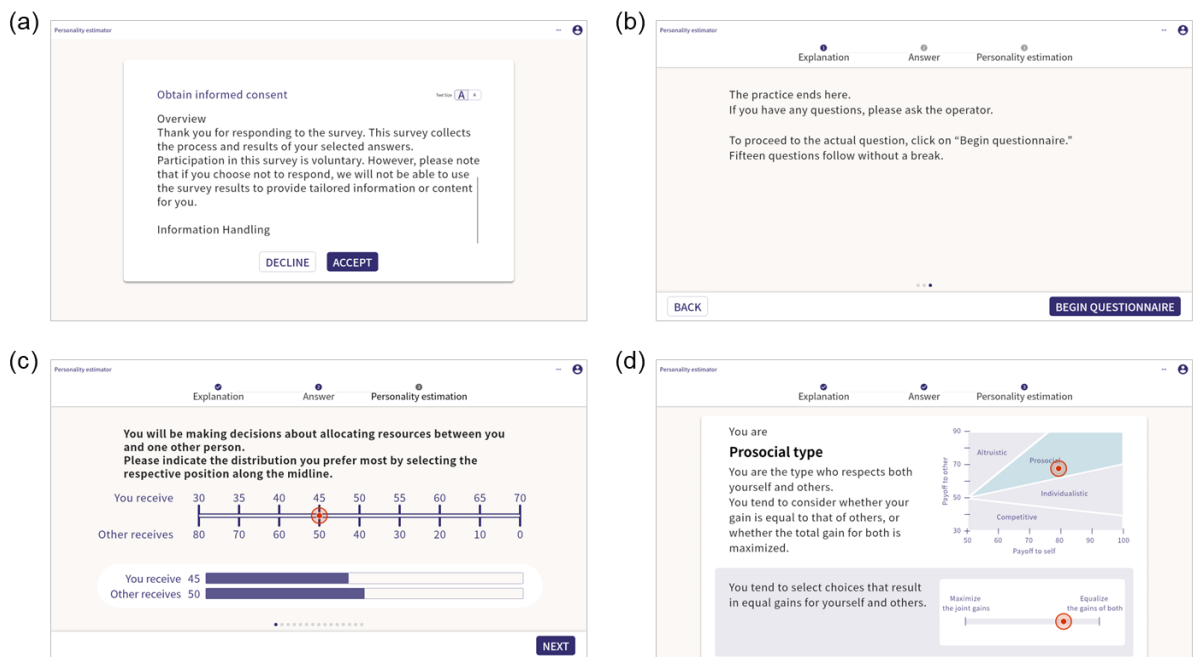

**FIGURE 13.** Personality estimator: User interface screens.

Figure 13a displays the consent acquisition screen as shown in Figure 6, where the policy operator obtains consent from the subjects regarding the questionnaire.

Figure 13b presents the questionnaire start screen following the explanation and practice phase shown in Figure 6. Figure 13c shows the questionnaire response screen for the subject, who answers 15 questions related to the primary SVO and secondary SVO using a slider. For details on the question content, see Appendix A.5.

Figure 13d displays the personality estimation screen as shown in Figure 6. The personality (primary SVO) is calculated using Equation (10), and the subject's personality is plotted on a two-dimensional graph divided into four regions (altruistic, prosocial, individualistic, competitive) at the upper part of the screen, along with an explanatory note for the subject. Additionally, the subject's equality orientation (secondary SVO) is shown as a slider at the lower part of the screen.

In the unused stock utilization project, to advance the (A x B) combined production and energy project utilizing abandoned farmland and neglected forests, it is preferable for subjects involved in both projects to exhibit altruistic or cooperative attributes. However, even when attributes are individualistic or competitive, cooperative behavior will be encouraged by the behavior change promoter discussed in the next subsection.

## 3.6 Behavior change promoter

Figure 14 shows representative screens of the behavior change promoter's user interface.

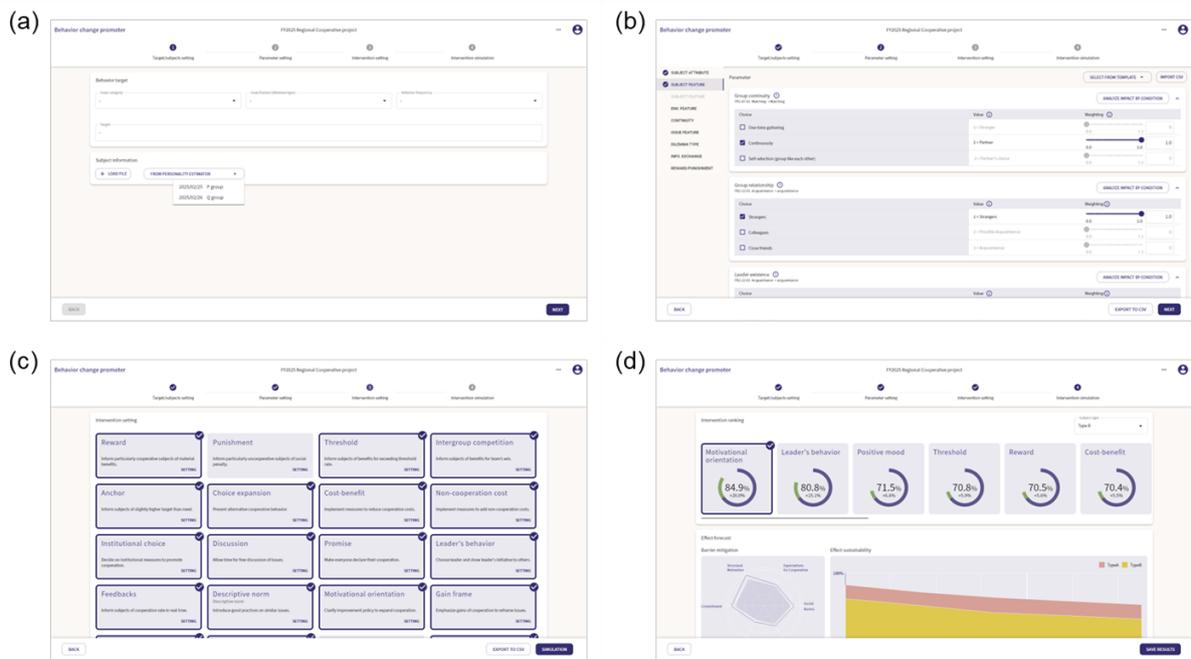

**FIGURE 14.** Behavior change promoter: User interface screens.

Figure 14a displays the behavior target setting screen (upper) and the subject setting screen (lower) as shown in Figure 7. In the former, issue categories, characteristics, and frequency of policy operation are specified as behavior targets. In the latter, subjects are either newly defined or imported from the personality estimator along with their attributes.

Figure 14b presents the parameter setting screen as shown in Figure 7. Initial parameters—such as the structure of the social dilemma, subject attributes, relationships among subjects, and the presence or absence of rewards and punishments—are configured as features of the social psychology model. At this stage, the composition of regional communities shown in Appendix A.1 is referenced. In this way, the psychological model described in Equation (11) is determined.

Figure 14c shows the intervention setting screen as shown in Figure 7. Multiple interventions from a predefined menu can be selected, or a new intervention can be created. By clicking on an intervention, the initial parameter set can be modified to increase the cooperation rate based on Equation (12). For details on the psychological model parameters, see Appendix A.6.

Figure 14d shows the intervention suggestion screen as shown in Figure 7. Interventions are displayed in ranked format at the upper part of the screen, listed in order of effectiveness for the subjects. When an intervention is selected, a radar chart illustrating improvements to the factors hindering cooperative behavior appears on the lower left, while the sustainability of the intervention for the subjects is shown on the lower right.

In Figure 14, in the unused stock utilization project of (A x B)—combined production and energy project utilizing abandoned farmland and neglected forests—the subjects are envisioned as owners of abandoned farmland and neglected forests with individualistic personalities. These owners are expected to exhibit cooperative behavior by making their abandoned farmland and neglected forests accessible to non-farming residents, immigrants, and private companies. Figure 14d indicates that motivational orientation, leader's behavior, and fostering a positive mood are effective interventions for these subjects. Implementing these interventions can encourage cooperative behavior and support the operation of policy (A × B).

## 3.7 Common mediator

Figure 15 shows representative screens of the common mediator's user interface, illustrating an avatar animation snapshot.

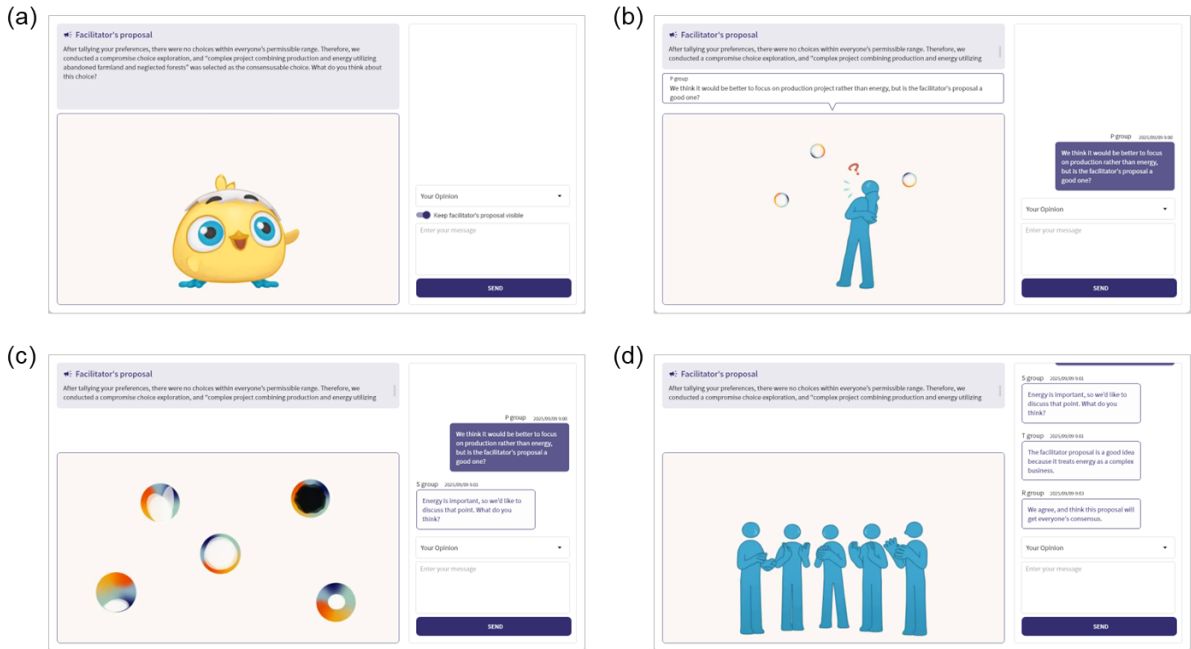

**FIGURE 15.** Common mediator: User interface screens.

Figure 15a displays the facilitator proposal screen in the consensus-building facilitator. It uses the chick type avatar shown in Figure 8c and facilitator motion #3 listed in Table 1. The right side of the screen serves as the chat space.

Figure 15b presents the divergent opinion screen for an individual participant. It combines the geometry type avatar in Figure 8a with the humanoid type avatar in Figure 8b, using individual participant motion #5 from Table 1.

Figure 15c shows the screen representing the scattered state of the participant group's field. It employs the geometry type avatar in Figure 8a and participant group motion #13 from Table 1.

Figure 15d displays the screen representing the state where the participant group's field has reached consensus. It uses the humanoid type avatar in Figure 8b and participant group motion #16 from Table 1.

As shown in Figure 15, incorporating avatar animation in the dialogue between the facilitator and deliberation participants enables smoother communication.

## 4. DISCUSSION

In this study, we have demonstrated that Social Co-OS supports human social systems based on a novel application architecture. Human society consists of three levels: individual behavior, interindividual interaction, and institutional formation. Social Co-OS performs collective intention diagnostics (social impact evaluator), prognostics (pluralistic policy simulator), and intervention (consensus-building facilitator) within the deliberative/political slow loop associated with institutional formation, while conducting individual behavioral diagnostics (personality estimator) and prognostics/intervention (behavior change promoter) within the operational/administrative fast loop associated with interindividual interaction. This study specifically presents the fundamental applications of a cyber system designed to cooperate with human society.

Furthermore, using an unused stock utilization project to address regional social issues as an example, we have demonstrated the usefulness of the Social Co-OS applications as follows: Subjectively, local farmers believe that policies such as (C) sales projects utilizing regional products from farmland and forests, including abandoned farmland and forests, and (B) regional energy projects utilizing abandoned farmland and neglected forests are effective. Objectively, (A) production projects utilizing abandoned farmland and neglected forests and (C) aforementioned sales projects exhibit relatively high social and environmental values. Among the policy choices that take these factors into account, the policy of (A × B)—a combined production and energy project utilizing abandoned farmland and neglected forests—can achieve consensus among diverse stakeholders. When operating the (A×B) policy, interventions such as motivational orientation, leader's behavior, and fostering a positive mood are effective for owners of abandoned farmland and neglected forests.

The Social Co-OS application developed in this study is available for a free trial in its beta version [40]. As a technical challenge, we aim to enhance usability and accessibility by obtaining user feedback through real-world verification. In the future, we will focus particularly on the following aspects:

- **Social Impact Evaluator:** Automatically generate objective inputs (policy elements) for the pluralistic policy simulator based on policies with high values in the sensitivity analysis results of the logic model.
- **Pluralistic Policy Simulator:** Automatically generate representative policy choices for the consensus-building facilitator (as a social choice planner, which was omitted in this study) based on policy elements with high values in the sensitivity analysis results of the multi-agent model.
- **Consensus Building Facilitator:** Incorporate AI-powered facilitation [41] and integrate it complementarily with citizen participation applications such as Decidim [42] and Liqlid [43], which provide discussion and voting.
- **Personality Estimator:** In addition to the questionnaire, incorporate methods such as emotion recognition and behavioral observation [44].
- **Behavior Change Promoter:** Improve the current psychological model based on social psychology experimental data by applying it to real-world problems and expanding the learning dataset [37].
- **Common Mediator:** Apply it to applications beyond the consensus-building facilitator and expand real-world use by incorporating multimodal approaches and leveraging AI and robots [45], extending beyond on-screen avatars.
- **User Interface:** Although this overlaps with the common mediator, it involves AI automatically generating the user interface in real time based on the reactions of deliberation participants or policy subjects.

Additionally, the common mediator and user interface in this study are limited to on-screen representations in the form of text and avatars; however, future developments could extend beyond the screen, for example, by giving them a physical embodiment such as a robot. Limiting the system to on-screen implementation is advantageous because it is software-based, imposes no restrictions on execution environments, and allows for easy updates. Nevertheless, from the perspective of enactivism [46], providing physical embodiment could actively enhance consensus building and behavior change through direct interaction with users. Since human perceive both verbal and non-verbal cues, physical embodiment is expected to facilitate intuitive understanding and foster responsible decision-making.

Note that the applications developed in this study can be widely used not only as components of Social Co-OS but also as independent application tools. For example, applications such as the social impact evaluator or pluralistic policy simulator can be applied to business improvement initiatives within local governments and corporations, assessments related to SDGs (Sustainable Development Goals) and ESG (Environmental, Social, and Governance) activities, and organizational growth and talent development through evaluation processes. The consensus-building facilitator can support decision-making in local governments and corporate organizations, workshops and collaborations between government and citizens, and initiatives for resident autonomy and community management. The personality estimator can be utilized in recruitment activities (e.g., employment decisions, aptitude assessments) and human resource development (e.g., transfers, career development) by local governments and companies. The behavior change promoter can address general social dilemmas such as waste disposal and environmental pollution, as well as engagement within corporate organizations. Finally, the common mediator can be applied across a wide range of use cases.

From a future perspective, in parallel with the release of the beta version, we plan to conduct field trials with local governments, cooperative corporations, and other stakeholders, and continue to improve the Social

Co-OS application based on user feedback. Moreover, by collaborating with startups and third parties offering complementary technologies, we aim to transform this into a broader co-operating platform between cyber and human societies. From the perspective of corporate management, the social impact evaluator can be linked to business visioning, the pluralistic policy simulator to business planning, the consensus-building facilitator to business decision-making, the personality estimator to human resource evaluation, and the behavior change promoter to organizational engagement [47]. By integrating these with an ERP (Enterprise Resource Planning) system, it becomes possible to evolve them into a cooperative management platform.

## 5. CONCLUSION

In this study, we designed a new application architecture for Social Co-OS as a CPS to facilitate cooperation between cyber and human societies, and implemented its user interface. Specifically, we developed application tools including a social impact evaluator, a pluralistic policy simulator, a consensus-building facilitator, a personality estimator, and a behavior change promoter, as well as a common mediator consisting of an avatar that serves as an interface between these tools and humans. Furthermore, we demonstrated the usefulness of these tools for policy co-making and policy co-operation by applying them to an example case: an unused stock utilization project, which represents a regional social issue.

  The Social Co-OS application enables democratic decision-making and collaborative operations. This is expected to promote the spread of digital democracy and cooperative platforms for citizens and workers, and to contribute to the emergence of cooperative management platforms aimed at shifting corporate governance from capitalists and executives to workers. In the future, Social Co-OS will support the development of human- and nature-oriented CPSs to reduce wealth inequality and resource waste, enhance well-being, restore the natural environment, and promote the transformation of the capitalist economy into an alternative model.

## AUTHOR CONTRIBUTIONS

**Takeshi Kato**: Conceptualization, Funding Acquisition, Methodology, Project Administration, Supervision, Visualization, Writing – Original Draft Preparation, Writing – Review & Editing.
**Misa Owa**: Data Curation, Formal Analysis, Investigation, Methodology, Resources, Software, Validation, Visualization, Writing – Original Draft Preparation.
**Jyunichi Miyakoshi**: Data Curation, Formal Analysis, Investigation, Methodology, Resources, Software, Validation, Visualization, Writing – Original Draft Preparation.
**Yasuhiro Asa**: Data Curation, Formal Analysis, Investigation, Methodology, Resources, Software, Validation, Visualization, Writing – Original Draft Preparation.
**Takashi Numata**: Data Curation, Formal Analysis, Investigation, Methodology, Resources, Software, Validation, Visualization, Writing – Original Draft Preparation.
**Yasuyuki Kudo**: Data Curation, Formal Analysis, Investigation, Methodology, Resources, Software, Validation, Visualization, Writing – Original Draft Preparation.
**Tadayuki Matsumura**: Data Curation, Formal Analysis, Investigation, Methodology, Resources, Software, Validation, Visualization, Writing – Original Draft Preparation.
**Kanako Esaki**: Methodology, Writing – Original Draft Preparation.
**Ryuji Mine**: Conceptualization, Funding Acquisition, Methodology, Project Administration, Supervision, Writing – Review & Editing.
**Toru Yasumura**: Data Curation, Methodology, Software, Validation, Visualization.
**Hikaru Matsunaka**: Data Curation, Methodology, Software, Validation, Visualization.
**Satomi Hori**: Conceptualization, Data Curation, Funding Acquisition, Methodology, Project Administration, Software, Supervision, Validation, Visualization, Writing – Original Draft Preparation.

## ACKNOWLEDGEMENTS

Professor Yasuo Deguchi of Kyoto University and the Kyoto Institute of Philosophy provided guidance on the paradigm shift toward Society 5.0 and on co-operation between cyber and human societies, while Emeritus Professor Yoshinori Hiroi of Kyoto University offered insights into community economics and policy-making. Corporate Chief Researcher Hiroyuki Mizuno of Hitachi Ltd. contributed to the conceptualization of Social Co-OS, while Chief Designer Takashi Shirasawa and Designer Miho Kobayashi of Hitachi Ltd. assisted with the user interface design. We extend our deepest gratitude to all of them.

## CONFLICT OF INTEREST

The authors declare no conflict of interest.

## FUNDING STATEMENT

The authors received no specific financial support for this study.

## DATA AVAILABILITY STATEMENT

The data that support the findings of this study are available from the corresponding author upon reasonable request.

## ORCID

Takeshi Kato         https://orcid.org/0000-0002-6744-8606
Misa Owa             https://orcid.org/0000-0002-2570-2354
Jyunichi Miyakoshi   https://orcid.org/0000-0003-2989-7619
Yasuhiro Asa         https://orcid.org/0000-0002-0349-6537
Takashi Numata       https://orcid.org/0000-0001-6981-1766
Yasuyuki Kudo        https://orcid.org/0000-0001-6476-8148
Tadayuki Matsumura   https://orcid.org/0009-0009-1907-1996
Kanako Esaki         https://orcid.org/0000-0002-3269-9130
Ryuji Mine           https://orcid.org/0000-0002-0130-6752

# APPENDIX

## A.1 Regional community composition

This study uses an unused stock utilization project in regional communities as an example. Based on the situation in Japanese rural areas [39], it envisions a regional community as follows:

- A relatively populous village located in mountainous rural areas (approximately 200 households and about 500 people).
- Within the village, approximately 60 farming and forestry households and about 140 non-farming and non-forestry households.
- Among non-agricultural households, approximately 40 are land-owning non-agricultural households.
- Approximately 50% of the farmland and forests owned by non-agricultural landowners are leased to farmer acquaintances, while the remaining 50% comprise abandoned farmland and neglected forests.
- One of the challenges of this project is to encourage non-agricultural landowners to make their abandoned farmland and neglected forests accessible to non-farming residents, immigrants, and private businesses who are outside their circle of acquaintances.

## A.2 Logic model

Generally, a logic model represents the relationship among inputs, activities, outputs, outcomes, and impacts using a logic tree. In this study, to examine the sensitivity of the impact (policy goal) to policy proposals, the first level of the logic model uses the policy proposal as the input; the next level uses direct outputs instead of activities, and short-term, medium-term, and long-term outcomes instead of outputs and outcomes [32]. Specifically, the combination of inputs (resources) and activities in a general logic model is treated as inputs (policy proposals) in this study.

In the unused stock utilization project in regional communities, as shown in Figure A1, the inputs consist of four policy proposals for utilizing abandoned farmland, neglected forests, and vacant homes, while the impact is defined as "realizing a sustainable regional community that coexists with nature, preserves culture, and achieves economic self-reliance." The blue and red arrows represent positive and negative relationships, respectively, and are assigned subjective weights.

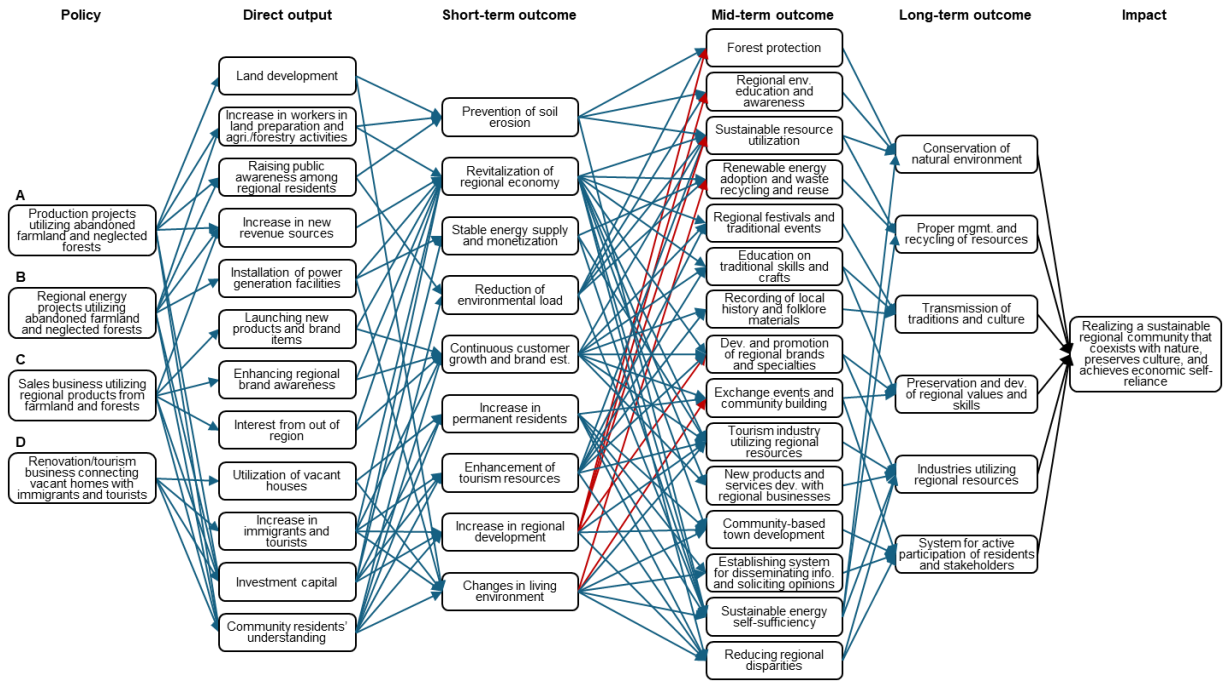

**FIGURE A1.** Logic model for unused-stock utilization policy project.

## A.3 Multi-agent model

Multi-agent models consist of inputs (policy elements), intermediate outcomes, and values related to society, the environment, and the economy [33]. In this study, as shown in Figure A2, funds are defined as the input; business, revenue distribution, and value factors are treated as intermediate outcomes; and these are linked to the final stage that represents social, environmental, and economic values. The blue and red arrows indicate positive and negative relationships, respectively, with weights assigned in the range of -1 to +1.

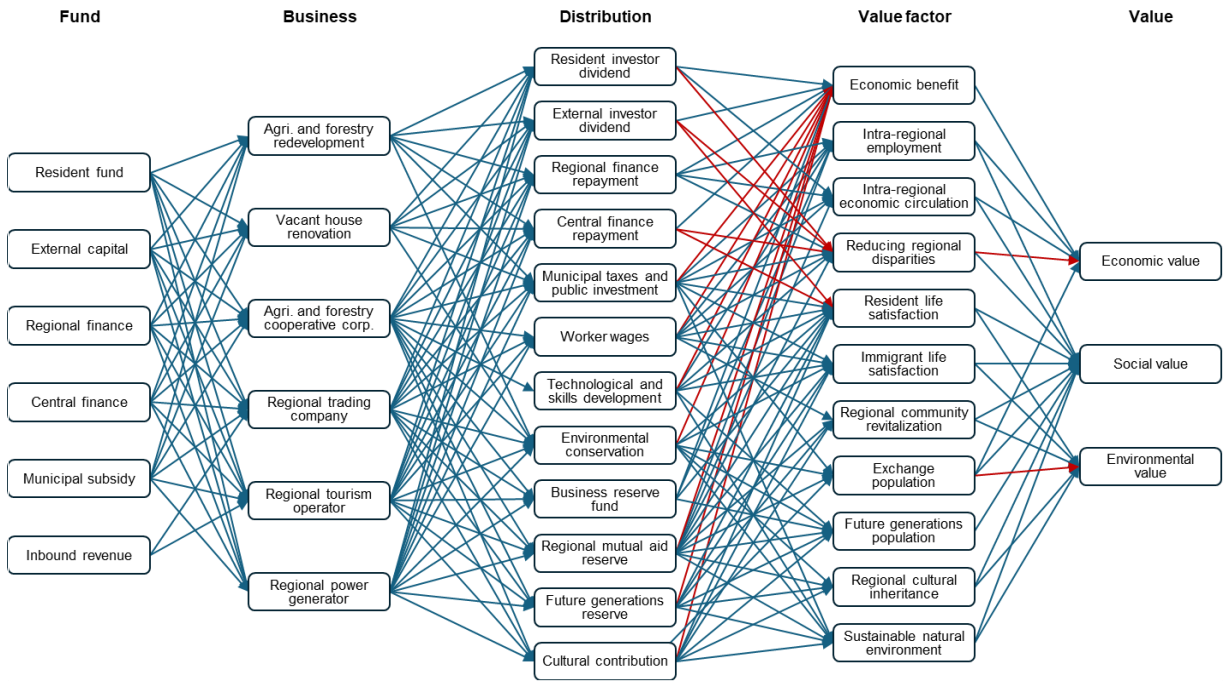

**FIGURE A2.** Multi-agent model for unused stock utilization policy project.

In the multi-agent model, each policy in the unused stock utilization project is represented by a combination of inputs, i.e., the allocation of funds. Table A1 provides examples of inputs corresponding to the four policy proposals shown in Figure A1.

TABLE A1. Inputs of multi-agent model for policy proposals.

| # | Input | Policy A<br>Production projects utilizing abandoned farmland and neglected forests | Policy B<br>Regional energy projects utilizing abandoned farmland and neglected forests | Policy C<br>Sales business utilizing regional products from farmland and forests | Policy D<br>Renovation/tourism business connecting vacant homes with immigrants and tourists |
|---|---|---|---|---|---|
| 1 | Resident fund | 0.1 | 0.1 | 0.1 | 0.1 |
| 2 | External capital | 0 | 0.5 | 0 | 0.1 |
| 3 | Regional finance | 0.7 | 0.1 | 0.2 | 0.2 |
| 4 | Central finance | 0 | 0.2 | 0 | 0.2 |
| 5 | Municipal subsidy | 0.2 | 0.1 | 0.7 | 0.1 |
| 6 | Inbound revenue | 0 | 0 | 0 | 0.3 |

## A.4 Policy choices and factors

Based on the results of the social impact evaluator and the pluralistic policy simulator, the consensus-building facilitator identified six policy choices, including (A) production project, (B) energy project, and (C) sales project, as well as combinations of these, as shown in Table A2. Additionally, in the sublated choice creation process, ten factors that constitute the policy choices are specified, as shown in Table A3.

TABLE A2. Policy choice options.

| Item | Policy | Details |
|---|---|---|
| A | Production projects utilizing abandoned farmland and neglected forests | Abandoned farmland and neglected forests are redeveloped to produce crops, wild vegetables, mushrooms, and other products. |
| B | Regional energy projects utilizing abandoned farmland and neglected forests | Locally sourced and consumed energy is produced using solar power and biomass, derived from from crop residues and thinned wood. |
| C | Sales business utilizing regional products from farmland and forests | Regional products from farmland and forests, including abandoned land, are sold at direct sales outlets and through e-commerce. |
| A×B | Complex project combining production and energy utilizing abandoned farmland and neglected forests | While producing crops on farmland, biomass energy is generated from crop residues and thinned wood and utilized . |
| B×C | Ecological sales business utilizing regional products and energy from abandoned farmland and forests | Utilizing regional energy sources, the direct sales outlet operates by selling regional products and promoting eco-friendly experiential tourism. |
| C×A | Sixth industry of local production for local consumption utilizing abandoned farmland and forests | Utilizing farmland and forests for production, integrating processing and sales to achieve sixth-sector industrialization. |

TABLE A3. Policy factors.

| Factor | Details |
|---|---|
| Revitalization of regional agriculture | Reconstruct agriculture as a sustainable and attractive industry by regenerating abandoned farmland, training young farmers, and introducing crops suited to the region. |

| Revitalization of regional forestry | Develop forestry as an industry that sustainably utilizes forest resources through the maintenance and thinning of neglected forests, the production, processing and utilization of timber, and the training of forestry workers. |
|---|---|
| Revitalization of tourism industry | Generate regional consumption by utilizing local resources (such as nature, agricultural and forestry experiences, landscapes, and culture) to attract visitors from outside the region. |
| Reduction in energy costs | Reduce energy expenditures by introducing renewable energy sources (such as biomass, solar power, and geothermal energy) to promote regional energy supply and self-sufficiency. |
| Environmental conservation | Protect the environment by maintaining ecosystems, combating climate change, and reducing disaster risk through sound management of land, water, and forests. |
| Job creation | Create employment opportunities for regional residents and immigrants through new production activities, service provision, processing and sales, tourism, and other means. |
| Business profitability | Establish a business model that is profitable and generates returns through business measures. |
| Establishment of regional brands | Enhance regional value by branding region-specific resources, products, and culture, and promoting them outside the region to gain recognition. |
| Intra-regional economic circulation | Strengthen mechanisms that circulate business revenue within the region (e.g., local production → local consumption, local employment → local spending). |
| Promotion of migration and settlement | Develop an environment that makes people want to stay—including housing, jobs, and community—to attract and retain new immigrants from outside the region and prevent residents from leaving the region. |

### A.5  Social value orientation questionnaire

The measurement of Social Value Orientation (SVO) is based on Reference [36]. Figures A3 and A4 show the slider scales for measurement used for the primary SVO and secondary SVO questionnaires, respectively.

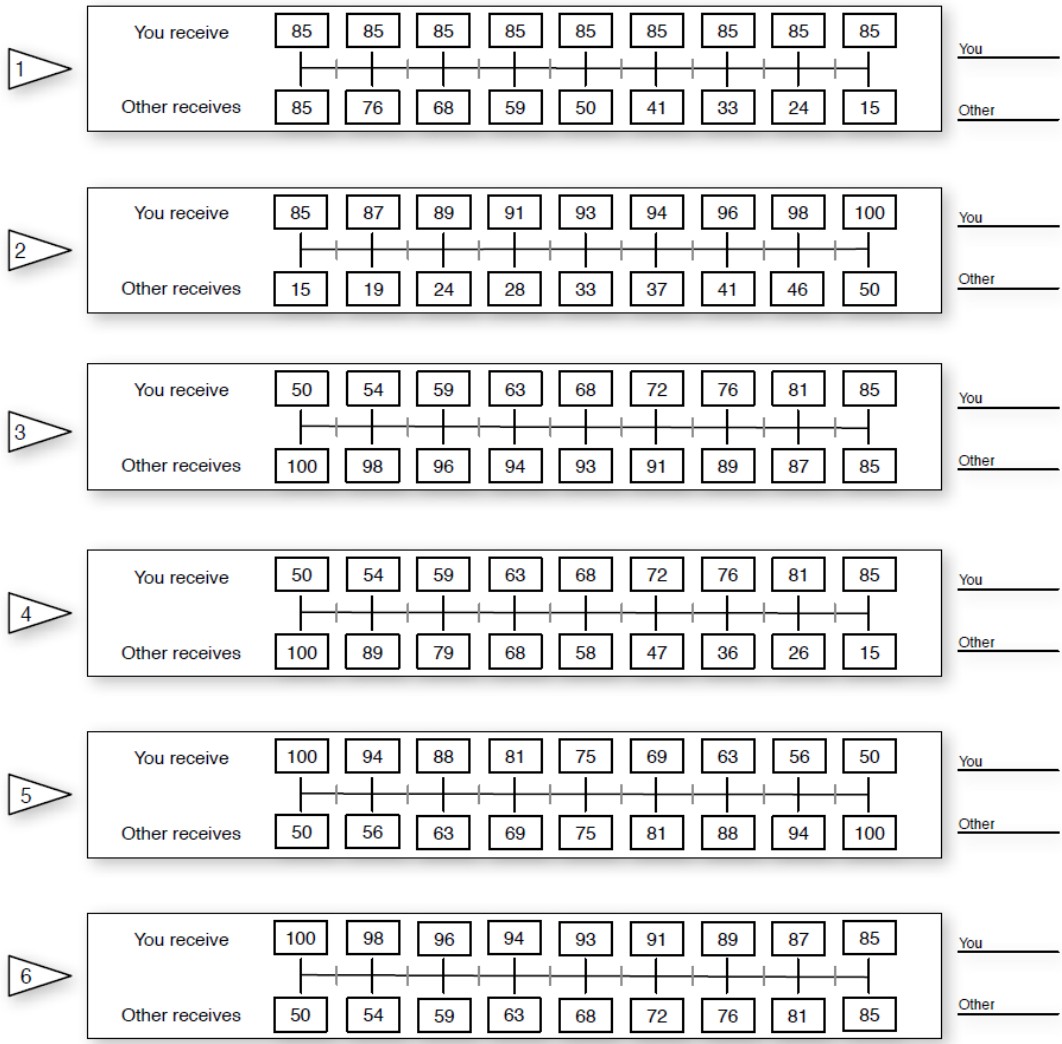

**FIGURE A3.** Primary SVO slider. (R. O. Murphy, K. A. Ackermann and M. J. J. Mandgraaf, 2011 [36], CC BY 3.0 License)

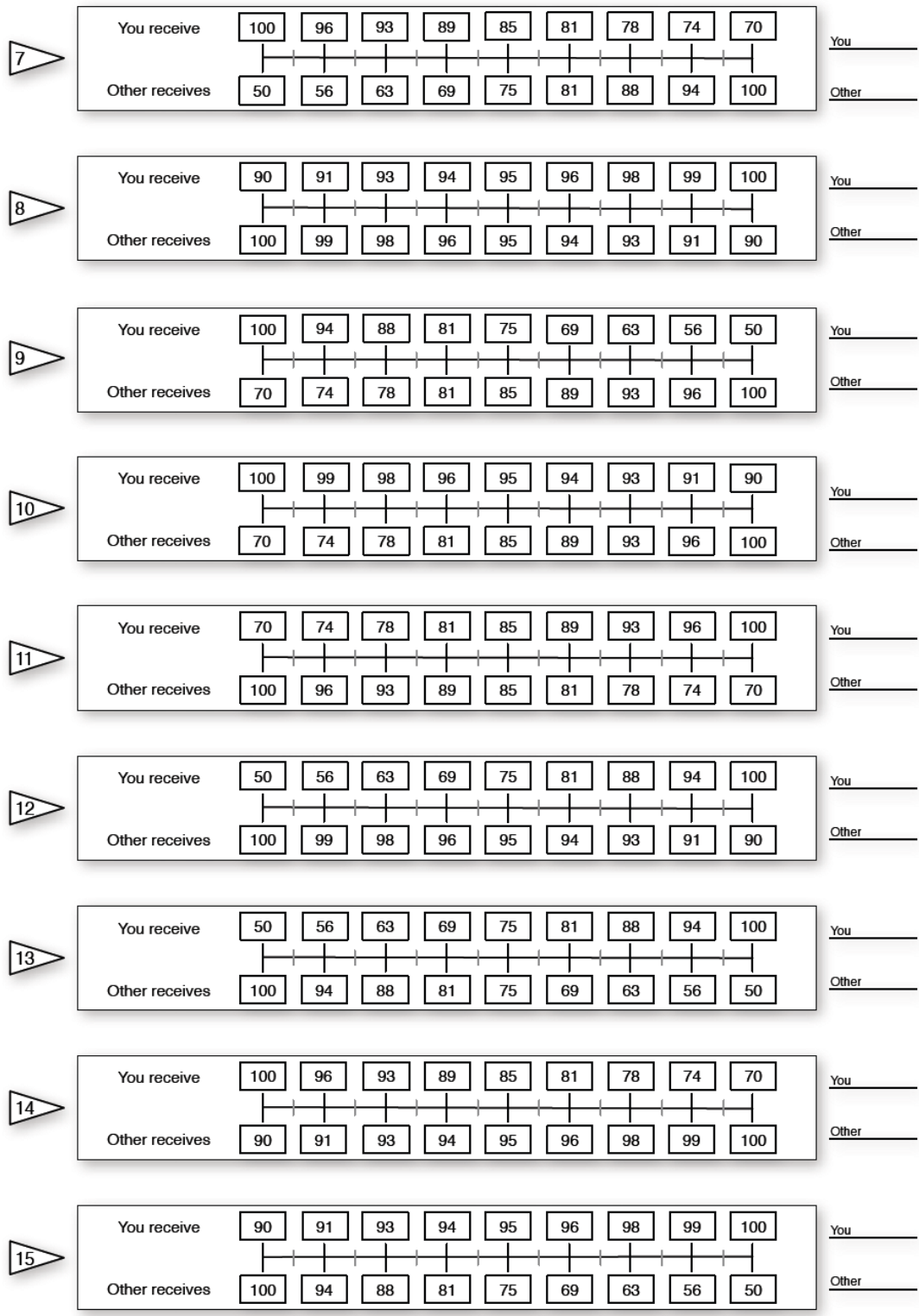

**FIGURE A4.** Secondary SVO slider. (R. O. Murphy, K. A. Ackermann and M. J. J. Mandgraaf, 2011 [36], CC BY 3.0 License)

## A.6 Psychological model parameter

The psychological model and its parameters are based on Reference [37], and Table A4 lists the parameters. Although numerous parameters exist, the behavior change promoter simplifies the user's parameter-setting process by preselecting representative parameters for policy operation.

**TABLE A4.** Psychological model parameters.

| # | Category | Parameter name | |
|---|---|---|---|
| 1 | Study Characteristics | Anchor | Anchor |
| 2 | | Choices | Number of choices |
| 3 | | | Lowest choice option |
| 4 | | | Highest choice option |
| 5 | | Exit option | Exit option |
| 6 | | Cognitive Load | Cognitive load treatment |
| 7 | | Discussion | Discussion |
| 8 | | | Communication type |
| 9 | | | Communication occurrence |
| 10 | | | Restricted communication |
| 11 | | | Communication content |
| 12 | | | Optional communication |
| 13 | | | Real communication |
| 14 | | Criticality | Criticality level |
| 15 | | Degree of conflicting interests | K index |
| 16 | | Conflict index | Conflict index |
| 17 | | | Conflict index value |
| 18 | | Experimental Setting | Experimental setting |
| 19 | | Feedback | Feedback |
| 20 | | | Feedback content |
| 21 | | Gain-Loss frame | Gain-Loss frame |
| 22 | | Give-Take frame | Give-Take frame |
| 23 | | Public Good-Bad frame | Public Good-Bad frame |
| 24 | | Focal point frame | Focal point frame |
| 25 | | Context frame | Context frame |
| 26 | | Number of trials | Number of trials |
| 27 | | Number of blocks | Number of blocks |
| 28 | | Number of sessions | Number of sessions |
| 29 | | One-shot vs repeated | One-shot vs repeated |
| 30 | | Known endgame | Known endgame |
| 31 | | One-shot vs repeated | Repeated one-shot game |
| 32 | | Continuation probability | Continuation probability |
| 33 | | Game type | Game type |
| 34 | | Behavior in different game | Behavior in different game |
| 35 | | Group size | Group size |
| 36 | | Decision maker | Decision maker |
| 37 | | | Size of unit of decision maker |
| 38 | | Group size multilevel ratio | Group size multilevel ratio |
| 39 | | Hormones administration | Hormones administration |
| 40 | | Hormones | Hormone |
| 41 | | | Hormone level |
| 42 | | Identification | Partner's group membership |
| 43 | | | Identification level |
| 44 | | Group type | Group type |
| 45 | | Knowledge of group membership | Knowledge of group membership |
| 46 | | Entitativity | Entitativity level |

| # | Category | Parameter name | |
|---|---|---|---|
| 47 | | Game incentive | Game incentive |
| 48 | | Show-up fee | Show-up fee |
| 49 | | Intergroup competition | Intergroup competition |
| 50 | | Institution | Rebate vs refund |
| 51 | | Institutional Choice | Institutional choice |
| 52 | | | Institutional choice mechanism |
| 53 | | | Institution type |
| 54 | | | Vote target |
| 55 | | | Vote outcome |
| 56 | | Leadership | Leadership |
| 57 | | | Leadership role |
| 58 | | | Endogenous leadership |
| 59 | | | Leadership assignment rule |
| 60 | | | Leader's behavior |
| 61 | | | Leader's characteristic |
| 62 | | Matching | Matching |
| 63 | | Partner choice | Partner choice |
| 64 | | Monitoring | Costly monitoring |
| 65 | | | Monitoring cost |
| 66 | | Motivational Orientation | Motivational orientation |
| 67 | | Endogenous motivational orientation | Endogenous motivational orientation |
| 68 | | Noise | Noise |
| 69 | | Descriptive norm | Descriptive norm |
| 70 | | Group variability | Group variability |
| 71 | | Ostracism | Ostracism |
| 72 | | Partner type | Partner type |
| 73 | | Iterated strategy | Iterated strategy |
| 74 | | | Iterated pre-programmed coop. rate |
| 75 | | One-shot strategy | One-shot strategy value |
| 76 | | | One-shot strategy |
| 77 | | Trial of cooperation | Trial of cooperation |
| 78 | | Trial of cooperation (ordinal) | Trial on which the DV was assessed |
| 79 | | Trial of cooperation (ordinal) | Block of cooperation |
| 80 | | Block of cooperation (ordinal) | Block of cooperation (ordinal) |
| 81 | | Physical proximity | Physical proximity |
| 82 | | Power | Power level |
| 83 | | | Power type |
| 84 | | | Power manipulation method |
| 85 | | Priming | Primed construct |
| 86 | | Endowment size | Endowment size |
| 87 | | Symmetric endowment | Symmetric endowment |
| 88 | | Assigned endowment | Assigned endowment |
| 89 | | Private account return level | Private account return level |
| 90 | | Symmetric private account return | Symmetric private account return |
| 91 | | MPCR | MPCR |
| 92 | | Individual MPCR | Individual MPCR |
| 93 | | External MPCR | External MPCR |
| 94 | | MRS | MRS |

| #   | Category | Parameter name |   |
|-----|----------|----------------|---|
| 95  |          | Symmetric MPCR | Symmetric MPCR |
| 96  |          | Threshold | Threshold |
| 97  |          |  | Continuous vs step-level public goods |
| 98  |          |  | Step return level |
| 99  |          |  | Endogenous threshold |
| 100 |          | Punishment | Punishment treatment |
| 101 |          |  | Punishment agent |
| 102 |          |  | Punishment incentive |
| 103 |          |  | Lottery punishment incentive |
| 104 |          |  | Sequential punishment |
| 105 |          |  | Punishment effectiveness |
| 106 |          |  | Punishment probability |
| 107 |          |  | Punishment rule |
| 108 |          |  | Punishment distribution rule |
| 109 |          |  | Punishment iterations |
| 110 |          |  | Focal participant has punished |
| 111 |          | Real partner | Real partner |
| 112 |          | Gossip | Gossip |
| 113 |          | Knowledge of partner's prior behavior | Knowledge of partner's prior behavior |
| 114 |          | Partner selection | Partner selection |
| 115 |          | Anonymity | Anonymity manipulation |
| 116 |          | Resource size | Resource size |
| 117 |          | Fixed resource | Fixed resource |
| 118 |          | Partitioned resource | Partitioned resource |
| 119 |          | Replenishment rate | Replenishment rate |
| 120 |          | Re-Start effect | Restart |
| 121 |          | Reward | Reward treatment |
| 122 |          |  | Reward agent |
| 123 |          |  | Reward incentive |
| 124 |          |  | Lottery reward incentive |
| 125 |          |  | Sequential reward |
| 126 |          |  | Reward effectiveness |
| 127 |          |  | Reward probability |
| 128 |          |  | Reward rule |
| 129 |          |  | Reward iterations |
| 130 |          |  | Focal participant has rewarded |
| 131 |          | Sequentiality | Sequentiality |
| 132 |          |  | Position in game |
| 133 |          |  | Endogenous position |
| 134 |          | Symmetry | Symmetry |
| 135 |          |  | Symmetry target |
| 136 |          | Synchrony | Synchrony |
| 137 |          | Taxation | Minimum contribution value |
| 138 |          |  | Minimum contribution |
| 139 |          | Income tax | Income tax |
| 140 |          |  | Income tax rate |
| 141 |          | Time pressure | Time pressure |
| 142 |          | Decision time | Decision time |

| # | Category | Parameter name | |
|---|---|---|---|
| 143 | | Uncertainty | Uncertainty target |
| 144 | | | Uncertainty level |
| 145 | | Watching eyes | Watching eyes |
| 146 | Sample Characteristics | Acquaintance | Acquaintance |
| 147 | | Age | Mean age |
| 148 | | | Lowest age |
| 149 | | | Highest age |
| 150 | | Age cohort | Age cohort |
| 151 | | Participant's own behavior | Participant's behavior level |
| 152 | | Partner's behavior | Partner's behavior level |
| 153 | | Education | Student sample |
| 154 | | | Academic discipline |
| 155 | | | Academic grade |
| 156 | | | Academic grade level |
| 157 | | | Academic performance |
| 158 | | Knowledge of experimental games | Knowledge of experimental games |
| 159 | | Emotions | Emotion |
| 160 | | | Emotion manipulation |
| 161 | | | Emotion valence |
| 162 | | | Emotion intensity |
| 163 | | Partner's emotion | Partner's emotion display |
| 164 | | | Partner's emotion |
| 165 | | | Partner's emotion manipulation |
| 166 | | | Partner's emotion valence |
| 167 | | Country or region | Country or region |
| 168 | | | Ethnicity (US) |
| 169 | | Source of country or region | Source of country or region |
| 170 | | Small-scale society | Heterogeneous ethnicity |
| 171 | | Expectations | Expectations level |
| 172 | | Comprehension of the game | Comprehension of the game level |
| 173 | | Proportion of males | Proportion of males |
| 174 | | Gender | Gender |
| 175 | | Partner's gender is known | Partner's gender is known |
| 176 | | Gender role | Gender role |
| 177 | | SVO type | SVO type |
| 178 | | Individual differences | Individual difference |
| 179 | | | Individual difference level |
| 180 | | Psychopathology | Psychopathology |
| 181 | | Partner perception | Partner perception |
| 182 | | | Partner perception level |
| 183 | | Similarity level | Similarity level |
| 184 | | Gender of the partner | Gender of the partner |
| 185 | | Preference for conditional cooperation | Preference for conditional cooperation |
| 186 | | Recruitment method | Recruitment method |
| 187 | | Religiosity | Religious exposure level |
| 188 | | | Religiosity operationalization |
| 189 | | Shadow of the future | Shadow of the future |
| 190 | | Social capital | Social capital level |

| #   | Category | Parameter name                  |                                 |
| --- | -------- | ------------------------------- | ------------------------------- |
| 191 |          | State trust                     | State trust level               |
| 192 |          | Values                          | Values (Schwartz)               |
| 193 |          | Year of data collection         | Year of data collection         |
| 194 |          | Source of year of data collection | Source of year of data collection |